\documentclass{article}

\usepackage[english]{babel}
\usepackage{natbib}
\usepackage[letterpaper,top=2cm,bottom=2cm,left=3cm,right=3cm,marginparwidth=1.75cm]{geometry}
\usepackage{placeins}
\usepackage{amsmath}
\usepackage{amsfonts}
\usepackage{float}
\usepackage{graphicx}
\usepackage{authblk}
\usepackage[colorlinks=true, allcolors=blue]{hyperref}
\usepackage{booktabs} 
\usepackage{tikz}
\usepackage{xcolor}
\usepackage{subcaption}
\title{Cities cluster into growth regimes that propagate shocks}

\author{Isaak Mengesha}
\author{Cillian Hourican}
\author{Debraj Roy}
\affil{Computational Science Lab, University of Amsterdam}
\begin{document}
\maketitle

\begin{abstract}
Economic growth is conventionally analyzed at the national level, yet cities generate the bulk of global output. Here we construct GDP trajectories for 8,808 functional urban areas (FUAs) across 165 countries over 1993–2019 using satellite-derived nighttime light data and identify 17 distinct, persistent growth regimes through clustering of full temporal trajectories. Cities within the same country frequently belong to different regimes, while structurally similar cities on different continents share the same one; regime membership explains 16\% of within-country growth variance beyond country fixed effects. A directed propagation network reveals that shocks transmit along lines of structural similarity rather than geographic proximity, with advanced economies exporting disturbances and emerging economies absorbing or amplifying them. The global economy might be better understood as an ecology of heterogeneous urban economies than as a collection of nations on a shared development path.
%Rather than converging toward a common frontier, FUAs inhabit distinct economic niches—analogous to ecological niches—defined by shared volatility profiles, shock responses, and long-run dynamics that transcend national boundaries. 

\end{abstract}

\section{Introduction}

Economic growth has long been analyzed primarily at the national level, with countries serving as the fundamental unit of observation in comparative development studies \cite{Barro1992, MankiwRomerWeil1992, Acemoglu2001}. This framing has yielded substantial insights into the mechanics of capital accumulation, technological diffusion, and institutional change \cite{Solow1956, Lucas1988, Acemoglu2005}. Yet national aggregation inevitably obscures critical heterogeneity in the spatial distribution of economic activity and its temporal dynamics \cite{Henderson1988, Desmet2015, Gennaioli2014}.
Cities and functional urban areas (FUAs) constitute the actual engines of growth, generating the bulk of global output through dense networks of production, innovation, and exchange \cite{Fujita1999, Glaeser2011}. A macro lens that treats countries as homogeneous units risks missing the fundamental organizational principles that govern how prosperity emerges, clusters, and propagates across space. The canonical neoclassical growth model \cite{Solow1956, Swan1956} posits that economies converge toward a common balanced growth path, differing only in initial conditions or structural parameters such as savings rates and population growth. Empirical tests of this framework have largely focused on cross-country regressions, finding evidence for conditional convergence among countries with similar fundamentals \cite{Barro1992, MankiwRomerWeil1992}. However, this convergence hypothesis rests on strong assumptions: diminishing returns to capital, exogenous technological change, and the existence of a single, globally shared technological frontier toward which all economies transition. Under this view, spatial patterns of development are incidental, and economic dynamics at sub-national scales are assumed to mirror those observed at the aggregate level.

Recent advances in satellite remote sensing and spatial data science have made it possible to move beyond country-level aggregates. Satellite data have been used to track structural transformations in urban form across thousands of cities \cite{frolking2024global} and to classify the temporal dynamics of urban activities over long historical periods \cite{gravier2024typology}. Nighttime lights (NTL) from the DMSP and VIIRS satellites are a well-tested proxy-measure for economic activity. They work best at telling apart which places are richer or poorer at a given moment: national-level fits give ($R^2\approx0.9$) \cite{Henderson2012, Chen2022}. They work less well at tracking how a single place changes year to year, and the lights-to-GDP relationship differs from place to place \cite{Gibson2021, Bluhm2022}. To mitigate this, we standardize each FUA's growth against all other FUAs in the same year, so the clustering compares where an FUA sits relative to its peers, not its absolute year-to-year growth. NTL also track city economies better than larger regional aggregates \cite{ChenNordhaus2011}, which suits our city-level (FUA) unit of analysis. A FUA is defined as cities and their surrounding commuting zones based on labor market integration \cite{BaumSnow2010,  Dijkstra2019, Florax2003,vermeulen2019modelling}.

In this paper, we leverage globally gridded GDP data for 8,808 FUAs across 165 countries over the period 1993-2019 to examine whether urban growth dynamics conform to the predictions of standard neoclassical models. We identify 17 distinct growth regimes by first reducing the dimensionality of growth trajectories via principal component analysis (PCA) and then partitioning the resulting representations using k-means clustering \cite{Lloyd1982, Arthur2007}. PCA preserves the dominant temporal structure of growth paths (e.g. boom-bust cycles, structural breaks, and volatility clustering) while removing noise-dominated dimensions, allowing us to group FUAs based on their characteristic dynamic signals. The choice for the number of clusters is determined by maximising the silhouette score \cite{Rousseeuw1987} over $k \in \{3, \ldots, 20\}$. The resulting clusters exhibit remarkable internal coherence and stability over time. 
%, suggesting that they are consistent with persistent attractor states in a multi-regime growth process rather than transient deviations from a single equilibrium.\\

We question the sensitivity of leading economic theories for growth and development with respect to scale. Given the substantial and increasing heterogeneity of prosperity within countries, it raises the question to what extent the national-level lens is suitable to assess their validity. Consider evolutionary economics, which posits that path dependencies in production and technological networks, based on historical contingencies, can lock economies onto differing paths \cite{Arthur1989, David1985, Dosi1982, Farmer2012, Arthur2013}. However, depending on how ``sticky'' human capital is to FUAs, it might be too strong an idealisation to simply assume frictionless markets within an entire nation, as the effectiveness of special economic zones among other things indicates. Whether or not these frictions (and consequently diverging trajectories) are caused by self-reinforcing capability accumulation \cite{Hidalgo2007, pinheiro2025dark} or by geographic contingencies such as spatial clustering and core-periphery dynamics goes beyond the scope of our investigation \cite{Krugman1991, Fujita1999, DurantonPuga2004}.

We document that subnational resolution reveals varying growth regimes, both across and within countries. For example, figure \ref{fig:cluster_regimes} shows strong differences across regimes, some exhibiting a single, others multiple dynamic attractors, whose locations span two orders of magnitude ($\sim$\$100 to $>$\$30k); a single pooled convergence slope therefore conflates within-regime catch-up with cross-regime divergence, and reads as one law only once regime membership is ignored. This heterogeneity operates within countries and is masked by national aggregates; regime membership explains 16\% of within-country growth variance beyond country fixed effects. Nor are these dependencies confined to national boundaries: shocks propagate between regimes along lines of structural similarity rather than geography, some regimes exporting disturbances and others absorbing them. This rich and diverging behaviour in the response to and production of shocks is further evidence for a more complex, network-based understanding of economic growth \cite{Starnini2019, Acemoglu2012}.

Taken together, at urban resolution a second organising layer becomes visible: an ecology of heterogeneous growth regimes that cuts across the national one. While this work attempts no causal identification, the meso-scale, border-crossing structure of these regimes is already enough to put the national unit of stabilization policy in question, and to invite closer investigation of the validity and effect sizes of the interdependencies operating on the urban systems of the globe.

\section{Results}
In this section we show that the growth regimes we recover are robust to alternative clustering methods and reasonable changes in the number of clusters. We then characterise their spatial structure and temporal dynamics, highlighting how structurally similar cities across countries share common trajectory types. Finally, we examine how shocks propagate between regimes through a sparse directed network, revealing distinct exporter, absorber, amplifier, and buffer roles in the global urban growth system.

\subsection{Robustness as prerequisite}

Interpreting ``growth regimes'' as economically meaningful units requires separating structure that is stable across reasonable modelling choices from patterns that are artefacts of a particular clustering pipeline. Any trajectory clustering exercise involves a chain of decisions—how to represent the time series, how to standardise them, how to reduce dimensionality, which algorithm to use, how many clusters $k$ to pick, and which internal validity metric to optimise—and none of these choices is uniquely ``correct'' \cite{jain2010data,kleinberg2003impossibility}. Internal criteria such as silhouettes or Jaccard coefficients encode assumptions about cluster geometry and can guide model selection, but they do not by themselves guarantee that the recovered regimes correspond to persistent economic structures rather than noise \cite{vonluxburg2012clustering,hennig2007clusterwise}. We therefore treat robustness as a prerequisite: regime-level claims must be supported by evidence that the underlying partition are robust to changes in the pipeline.

\begin{figure}[H]
    \centering
    \begin{subfigure}[t]{0.48\linewidth}
        \centering
        \includegraphics[width=\linewidth]{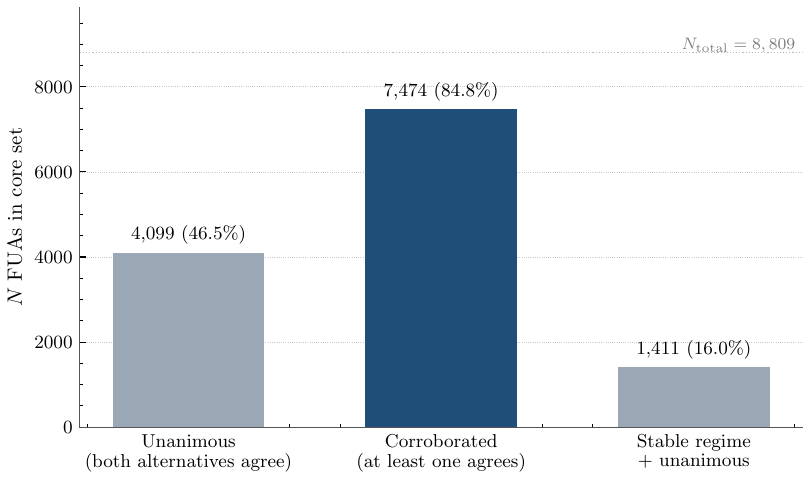}
        \caption{Core-set size under three method-robustness criteria.}
        \label{fig:abc_sizes_a}
    \end{subfigure}
    \hfill
    \begin{subfigure}[t]{0.48\linewidth}
        \centering
        \includegraphics[width=\linewidth]{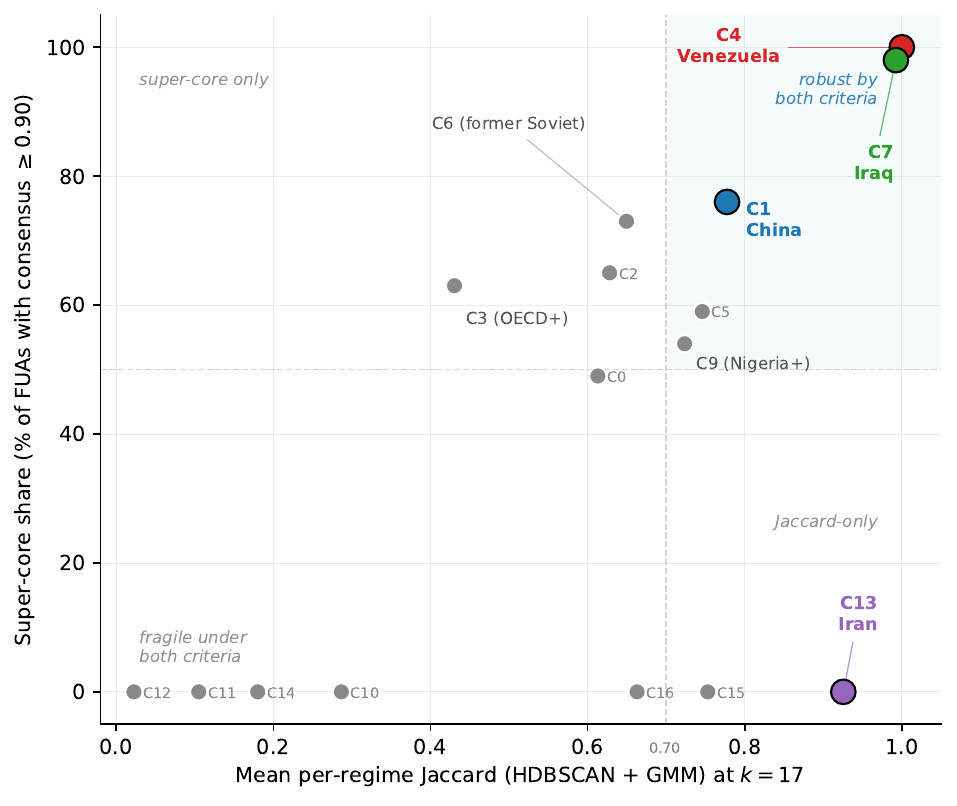}
        \caption{Two robustness criteria, jointly.}
        \label{fig:abc_sizes_b}
    \end{subfigure}
    \caption{\textbf{Regime robustness under two criteria.} \textbf{(a)}~FUAs retained under three method-robustness criteria (cross-sectional standardisation, $N_{\mathrm{total}} = 8{,}809$): unanimous (HDBSCAN and GMM both agree with k-means), $4{,}099$ ($46.5\%$, 14/17 regimes); corroborated (at least one agrees; used in main text), $7{,}474$ ($84.8\%$, all 17); stable-regime $+$ unanimous, $1{,}411$ ($16.0\%$; C1, C4, C7, C13). \textbf{(b)}~Mean per-regime Jaccard against HDBSCAN and GMM at $k = 17$ (Table~\ref{tab:jaccard}) vs.\ super-core share (fraction of FUAs with pairwise co-assignment $\geq 0.90$ across seven partitions; Supplementary~\ref{sec:supercore}). C1, C4, C7 are robust under both criteria; C13 passes Jaccard only (merges at $k = 18$); C6 super-core only. The criteria are complementary, not equivalent.}
    \label{fig:abc_sizes}
\end{figure}

We construct a ``core set'' of FUAs whose regime assignments are corroborated by independent clustering methods (Methods~4.5). Re-clustering the same PCA-reduced trajectories with a density-based algorithm (HDBSCAN) and a Gaussian mixture model (GMM), and matching labels to the $k$-means reference partition via the Hungarian algorithm, yields per-regime Jaccard indices reported in Supplementary Table~3. An FUA enters the core set if at least one of the two alternative methods assigns it to the same regime as $k$-means. This corroborated criterion retains $N_{\text{core}} = 7{,}474$ FUAs (84.8\% of the sample) spanning all 17 regimes (Figure~\ref{fig:abc_sizes}); these FUAs form the basis for all regime-level analyses in the main text. The remaining 1{,}334 FUAs sit on regime boundaries where algorithms disagree, and we exclude them when characterising regime behaviour. A stricter unanimity rule, which requires both HDBSCAN and GMM to agree with $k$-means, still retains 4{,}099 FUAs (46.5\%) across 14 of the 17 regimes, indicating that a substantial fraction of cities are grouped consistently regardless of algorithm.

The same robustness check applies at the level of entire regimes. We apply two complementary approaches: Jaccard tests whether the same regime is recovered at a single $k$; the super-core tests whether its members stay grouped together when both the method and $k$ are varied. At fixed $k = 17$, per-regime Jaccard indices against HDBSCAN and GMM identify four \emph{anchor regimes} whose memberships are recovered particularly consistently (Supplementary Table~\ref{tab:jaccard}): a China-dominated regime (C1), Venezuela (C4), Iraq (C7), and Iran (C13). A complementary super-core measure (Supplementary~\ref{sec:supercore}) confirms C1, C4, and C7 as robust under both criteria. Two regimes pass only one: C13 by Jaccard but not super-core (its members merge into a larger cluster at $k = 18$), and C6 (former-Soviet bloc) by super-core but not Jaccard. 
 Figure~\ref{fig:abc_sizes}b visualises both criteria  jointly; the anchor regimes correspond to distinctive combinations of industrialisation timing, conflict dynamics, resource dependence, and institutional regimes, suggesting that the clustering picks up persistent structural features rather than algorithm-specific artefacts.

Additionally, we investigate whether the choice of $k$ (i.e. the number of clusters) itself is fragile. Supplementary Figure~12 compares the $k=17$ partition similar values of $k$. When we apply $k$-means with $k=15$ or $k=20$, the resulting partitions mostly align with $k=17$ partitions instead of producing very novel partitions. Combined with adjusted Rand index comparisons across preprocessing variants reported in Supplementary Section~5.10, these results indicate that our main regimes, though containing fuzzy boundaries, contain robust and non-trivial underlying regimes.

\subsection{Spatial structure and regime behaviour}
\FloatBarrier

\begin{figure}[h]
    \centering
    \includegraphics[width=1\linewidth]{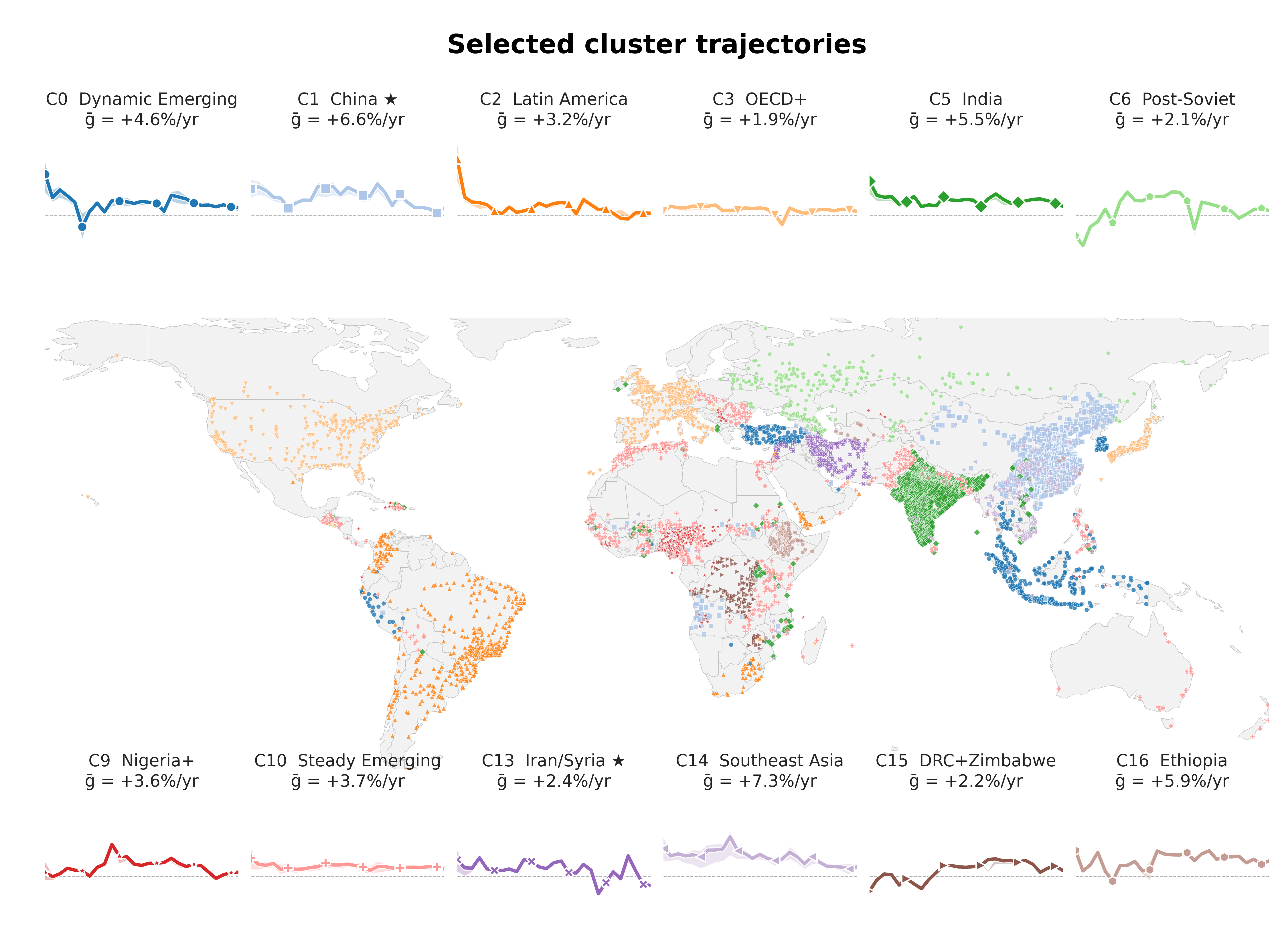}
    \caption{
        \textbf{Growth regimes are geographically coherent and temporally distinct without geographic input to clustering.}
        Spatial distribution of 8,808 FUAs (centre) coloured by regime membership from PCA--k-means on 1993--2019 growth trajectories, with mean growth trajectories $\bar{g}_{c,t}$ for the 12 main regimes shown above and below (annual mean $\bar{g}$ reported per panel; five small or single-country regimes omitted, see SI). Stars ($\star$) mark cross-method-stable regimes (Section~\ref{sec:robustness}). Regimes form coherent geographic blocs---former Soviet states in C6, China in C1, Latin America in C2---while also crossing national borders (Mexico clusters with the US and Canada in C3) and fragmenting large countries across multiple regimes. Volatility spans 28-fold across regimes, from 0.66\% (C10) to 18.93\% (C7, Iraq, omitted), reflecting genuine heterogeneity in shock exposure rather than measurement error.
    }
    \label{fig:clusteroverview}
\end{figure}

The clustering procedure (see Methods) is applied to time-series GDP growth data across all recoverable FUAs, excluding a small number such as Afghanistan. The recovered regimes are geographically coherent: although the input contains no geographic information, the clusters reproduce recognisable country groupings, with former-Soviet FUAs in C6, Chinese FUAs in C1, and most of Latin America in C2. Volatility varies sharply across regimes. C7 (Iraq), at 18.93\%, exceeds stable low-volatility regimes such as C3 (1.45\%) by more than an order of magnitude. This spread is larger than any plausible bound on NTL measurement noise, so the volatility differences in Figure~\ref{fig:regimetypology} reflect genuine heterogeneity in shock exposure rather than measurement artefacts.

Regime membership does not track income or development level. Single-country regimes such as C4 (Venezuela) and C7 (Iraq), with volatilities of 5.6\% and 18.9\%, are extreme-shock outliers where institutional breakdown overrides ordinary cyclical variation. Other regimes cut across development levels: C3 spans 42 nations and 1,208 FUAs, grouping OECD economies (US, Japan, Germany) with low-income countries (Haiti, Burundi, Madagascar). The unifying feature is trajectory shape, namely low volatility, stable trend, and synchronised cycle response. Income does not organise C3; institutional stability and structural integration do. We decompose the regimes by two summary statistics: its mean annual rate $\bar{g}_c$ and its within-regime volatility $\sigma_c$. More detailed behavioral descriptions, such as structural breaks, conditional dynamics, and asymmetric shock responses, can be speculated about but are not supported by the data used in this study. 

\begin{figure}[t]
\centering
\begin{tikzpicture}[scale=1]

\definecolor{maturecol}{HTML}{DCE6F1}
\definecolor{turbslowcol}{HTML}{F0DDC4}
\definecolor{steadycol}{HTML}{D8E8D0}
\definecolor{turbexpcol}{HTML}{F2D4D2}

\fill[maturecol]   (0,0) rectangle (5,3);
\fill[turbslowcol] (5,0) rectangle (10,3);
\fill[steadycol]   (0,3) rectangle (5,6);
\fill[turbexpcol]  (5,3) rectangle (10,6);

\draw[thick, black!55] (0,0) rectangle (10,6);
\draw[thick, black!55] (5,0) -- (5,6);
\draw[thick, black!55] (0,3) -- (10,3);

\node[align=center] at (2.5, 1.5) {%
    \textbf{Mature}\\[3pt]
    {\large C3}\\[5pt]
    \footnotesize\itshape OECD-anchored,\\ structurally integrated
};
\node[align=center] at (7.5, 1.5) {%
    \textbf{Turbulent slow growth}\\[3pt]
    {\large C2,\, C6,\, C13}\\[5pt]
    \footnotesize\itshape Latin America, post-Soviet,\\ Iran/Syria
};
\node[align=center] at (2.5, 4.5) {%
    \textbf{Steady expansion}\\[3pt]
    {\large C5,\, C14}\\[5pt]
    \footnotesize\itshape India, steady emerging,\\ W.\ Africa, SE Asia
};
\node[align=center] at (7.5, 4.5) {%
    \textbf{Turbulent expansion}\\[3pt]
    {\large C0,\, C1,\, C9}\\[5pt]
    \footnotesize\itshape Dynamic emerging, China,\\ Nigeria+, Ethiopia+
};

\node[rotate=90, anchor=south, font=\small\bfseries] at (-1.4, 3) 
    {Mean growth rate $\bar{g}$};
\node[anchor=east, font=\footnotesize] at (-0.15, 1.5) {$< 3.5\%$};
\node[anchor=east, font=\footnotesize] at (-0.15, 4.5) {$\geq 3.5\%$};

\node[anchor=north, font=\small\bfseries] at (5, -0.9) 
    {Within-regime volatility $\sigma_g$};
\node[anchor=north, font=\footnotesize] at (2.5, -0.1) {$< 2\%$};
\node[anchor=north, font=\footnotesize] at (7.5, -0.1) {$\geq 2\%$};

% \node[anchor=north west, align=left, font=\footnotesize, 
%       text width=10cm] at (0, -1.4) {%
%     \textbf{Excluded:} 
%     C4 (Venezuela), C7 (Iraq), C12 --- $\sigma_g > 5\%$, 
%     single-country institutional disruption. 
%     C15 (DRC, Zimbabwe) --- mean growth depends on sub-period chosen. 
%     These regimes are analysed separately in 
%     Section~\ref{sec:propagation}.
% };

\end{tikzpicture}
\caption{\textbf{Trajectory-shape typology of growth regimes.} The thirteen regimes outside the extreme-volatility band sort into  four quadrants defined by mean annual growth rate $\bar{g}$ and  within-regime volatility $\sigma_g$ (values from  Table~\ref{tab:cluster_stats}; thresholds set at $\bar{g}=3.5\%$  and $\sigma_g=2\%$, approximating the medians of the included  regimes). Representative regimes per quadrant. }
\label{fig:regimetypology}
\end{figure}
\

The thirteen included regimes sort into the four quadrants of Figure~\ref{fig:regimetypology}. With thresholds near the regime medians ($\bar{g} \approx 3.5\%$, $\sigma_g \approx 2\%$). The mature quadrant contains only C3, the OECD-anchored cluster. Estimated fixed points of the conditional response $\mathbb{E}[y_{t+1} \mid y_t]$ span two orders of magnitude across regimes ($\sim\!\$100$ in C9 to $>\!\$30{,}000$ in C3; Supplementary~\ref{fig:cluster_regimes}), and curvature varies in kind: concave saturation in C1, C13, C14; near-linear in C3, C10; multiple tentative fixed points in C9, C10, C15, C16. Saturation is sharpest in the cross-method-stable regimes (C1, C13). Distinct attractor locations are difficult to reconcile with a single global frontier and are the signature one would expect from regime-specific Solow parameters, path-dependent lock-in, or capability ceilings; distinguishing among these requires subnational structural data not currently available at FUA resolution. Note however that all regimes are in absolute terms growing. Meaning the macro level growth trap is not something we observe on the FUA level, instead these fixed points constitute relative growth. 

%\FloatBarrier

\subsection{Shock propagation and network structure}

\begin{figure}[H]
    \centering
    \includegraphics[width=1\linewidth]{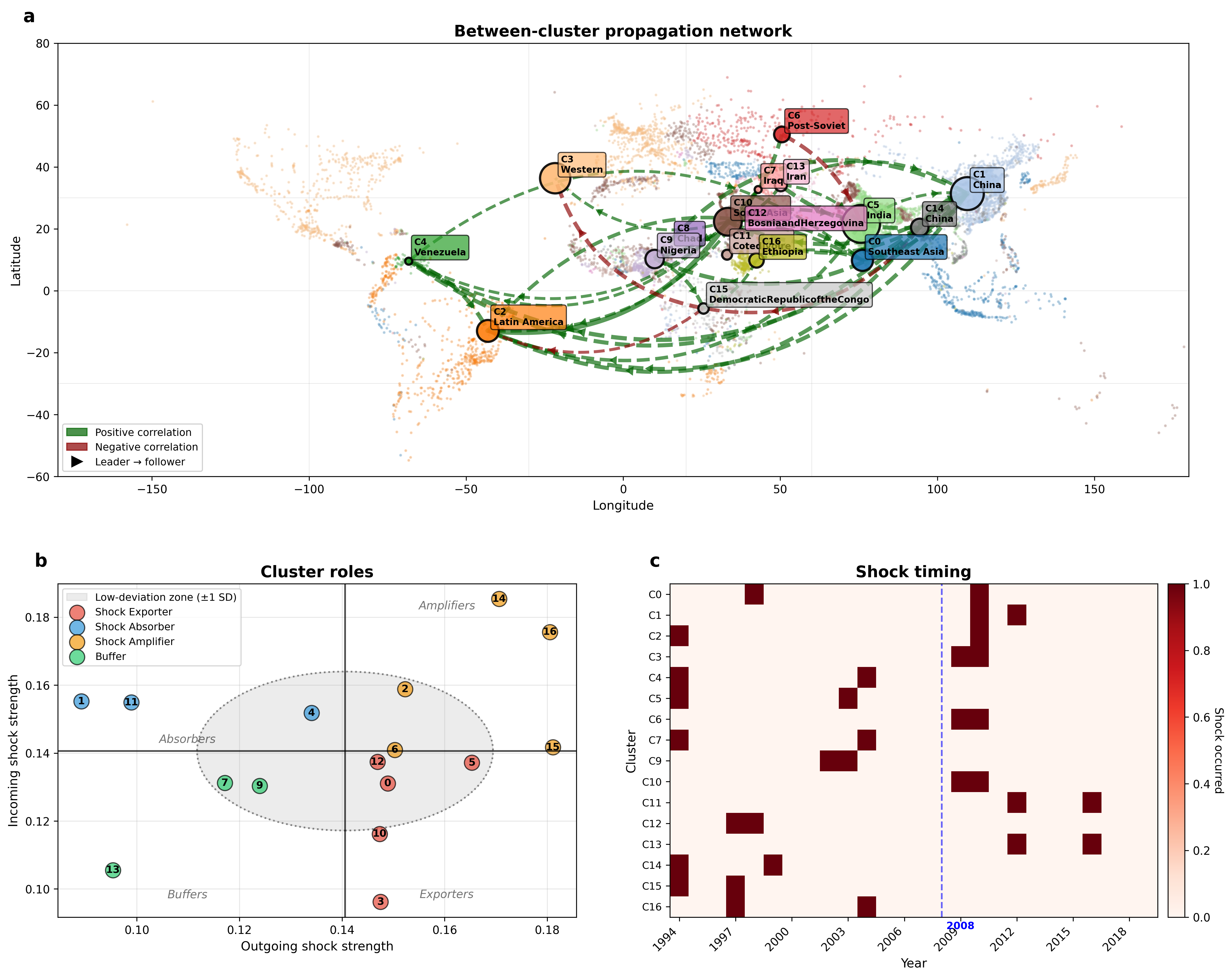}
    \caption{
    \textbf{Heterogeneous shocks propagate through a sparse directed network shaped by structural similarity, not geography.}
    \textbf{(a)} Between-cluster propagation network on the world map. Edges are statistically significant ($p \leq 0.05$) lagged correlations between cluster growth rates (lag 0--3 years); arrows point leader~$\to$~follower, green/red denote positive/negative correlations, width scales with strength. Node positions are geographic centroids, but edges cut across continents.
    \textbf{(b)} Cluster roles in the (outgoing, incoming) shock-strength plane, partitioned into amplifiers, absorbers, exporters, and buffers; shaded ellipse marks the $\pm 1$ SD low-deviation zone. Cluster~3 (advanced economies) is a net exporter; Clusters~1 (China) and 11 (West Africa) are net absorbers.
    \textbf{(c)} Shock years (top/bottom 2\% of each regime's growth distribution), 17 clusters, 1993--2019; dashed line marks 2008. Globally synchronised episodes (1997--1998, 2008--2009, 2014--2016) coexist with regime-specific shocks (Iraq 2003--2005, Zimbabwe 2000--2008), contradicting the common-shock assumption in standard growth regressions.
    Lagged correlations establish temporal precedence but not causal mechanism; co-exposure to common upstream shocks remains a possible alternative.
}
\label{fig:shocks}
\end{figure}

The lagged-correlation network (Figure~\ref{fig:shocks}) indicates that growth regimes are not isolated: significant directed co-movement links them along axes of structural similarity rather than geography. Each regime can be summarized by two scalars: total outgoing- and incoming- lagged correlation strength: Jointly they define four propagation roles (Table~\ref{fig:regimetypology}): \textit{exporters} (high out, low in), \textit{absorbers} (low out, high in), \textit{amplifiers} (high both), and \textit{buffers} (low both). These roles map onto the trajectory-shape quadrants of Figure \ref{fig:shocks} in a near-systematic way: the volatility axis ($\sigma_g$) separates amplifiers from buffers, while net direction is largely orthogonal to it. The mature quadrant (C3) supplies the only exporter; the turbulent quadrants (C0, C2, C6, C13, C16) and the extreme-volatility singletons (C4, C7, C12) populate the amplifier role; and the steady-expansion quadrant (C5, C10, C14) supplies most of the buffers. Two clusters break the pattern in informative ways: C1 (China-led, turbulent expansion) and C11 (West Africa, steady expansion) are absorbers---low- and high-volatility regimes can both end up receiving more than they transmit.\\

Three features of this network structure stand out. First, \textit{centrality need not imply connectivity}: C3 has the largest net export ($+0.051$) but is less densely linked than several emerging-market clusters. This might be an artifact of their overall stability, obfuscated by our correlation based approach. After all C3 combines the largest market power of all regimes. Second, \textit{economic size does not determine temporal position}: despite China's weight in global GDP, Cluster~1 absorbs ($-0.066$)---its fluctuations followed rather than led others over 1993--2019, plausibly reflecting the role of external demand and global financial conditions during the sample period. The classification captures temporal ordering, not mass. Third, \textit{integration is a double-edged sword}: buffer regimes contracted less in 2008--2009 than exporter or amplifier regimes, so the insulation conferred by low connectivity---whether driven by domestic demand (C5), weak financial integration (C14), or conflict-severed external links (C15)---also carries resilience. Buffer status is not necessarily desirable, but it can have benefits.\\

Two further observations qualify the network reading. Spatial decay of correlation strength with great-circle distance is weak and noisy (Figure~\ref{fig:spatialdecay}): many distant regime pairs co-move more strongly than proximate ones, reinforcing the structural-similarity interpretation. The shock-timing heatmap (Figure~\ref{fig:shocks}) shows that while a handful of events synchronize across regimes---the 1997--1998 Asian crisis, the 2008--2009 global financial crisis, the 2014--2016 commodity collapse---most shocks are regime-specific (Iraq 2003--2005, Zimbabwe 2000--2008, Russia 1998 with surprisingly limited spillovers). This heterogeneity is hard to reconcile with the common-shock process assumed by standard panel growth regressions. Throughout, we note that lagged correlations establish temporal precedence but cannot distinguish causal transmission from co-exposure to upstream shocks; identifying specific channels would require instrumental-variable or quasi-experimental designs beyond our setup.

\section{Discussion}

Taken together, we first confirmed the robustness of our regimes; distinguishing very stable \textit{anchor regimes} and reasonably stable more method dependent regimes.  Regimes exhibit distinct growth behaviour regarding fixed points, volatility and growth rate. Furthermore we found evidence, that they influence each other allowing us to estimate a directed network of shock propagations. We interpret this as evidence for a more complex "ecology of growth regimes", inviting further investigation.

\subsection{Convergence without a single frontier}

Our results are easily misread as a rejection of convergence. They are not. Convergence is among the most replicated regularities in growth empirics \cite{Barro1992, MankiwRomerWeil1992}, and we reproduce it: within regimes, the conditional growth response $\mathbb{E}[y_{t+1}\mid y_t]$ bends toward a regime-specific fixed point, so that poorer FUAs grow faster than richer ones in the same regime (althrough there are notable exceptions). This concavity is sharpest in the anchor regimes, where the partition is most robust (Supplementary Figure~\ref{fig:cluster_regimes}). What we reject is more specific: that a \emph{single} such process exists, common to all economies and oriented toward one global frontier. We separate three implications that the word ``convergence'' might obfuscate. The first (~\ref{fig:convergence_objects}a) is the unconditional $\beta$-convergence common in macroeconomic analysis, which operates globally and we compute on the FUA instead of the national level \cite{patel2021new}.
The second is cross-regime convergence (~\ref{fig:convergence_objects}b), which includes the same FUAs, partitioned by regime, each with idiosyncratic convergence behavior. The third is within-regime divergence; in case of multiple  (dynamic) fixed points. All regimes grow, but the differences in rates, result in some FUAs within regimes converging relative to one another while others diverge (~\ref{fig:convergence_objects}c). Many regimes have multiple other only a single attractor, spanning up to two order of magnitudes in per-capita GDP.

This is not to say economic dynamics between regimes differ more than they are similar; it does beg the question what predicts regime membership other than countries, and what explains the deviations from what would global $\beta$-convergence on the FUA level would predict otherwise.

\begin{figure}[H]
\centering
\begin{tikzpicture}[scale=0.85]

\definecolor{regA}{HTML}{A9C2DC}
\definecolor{regB}{HTML}{5E86B0}
\definecolor{regC}{HTML}{2C4A6B}
\definecolor{poolgrey}{HTML}{9AA0A6}

%% ================= PANEL (i): pooled regression =================
\begin{scope}[xshift=0cm]
  \draw[black!55,thick] (0,3.4) |- (4.6,0);
  \node[rotate=90,font=\scriptsize,black!60] at (-0.34,1.7) {growth};
  \node[font=\scriptsize,black!60] at (2.3,-0.44) {initial income};
  \node[font=\small\bfseries] at (2.3,4.05) {(a) Pooled regression};
  \node[font=\scriptsize\itshape,black!55] at (2.3,3.66) {one straight slope};
  \foreach \x/\y in {0.6/2.90,1.3/2.62,2.1/2.50,3.0/2.22,4.1/2.26,
                     0.7/2.30,1.5/1.50,2.2/1.00,3.1/1.60,4.0/2.00,
                     0.6/1.30,1.4/0.90,2.1/1.35,3.0/0.95,4.1/1.20}
     \fill[poolgrey] (\x,\y) circle (2.3pt);
  \draw[black,line width=1.4pt] (0.35,2.05) -- (4.45,1.45);
\end{scope}

%% ================= PANEL (ii): within regimes =================
\begin{scope}[xshift=5.9cm]
  \draw[black!55,thick] (0,3.4) |- (4.6,0);
  \node[rotate=90,font=\scriptsize,black!60] at (-0.34,1.7) {growth};
  \node[font=\scriptsize,black!60] at (2.3,-0.44) {initial income};
  \node[font=\small\bfseries] at (2.3,4.05) {(b) Within regimes};
  \node[font=\scriptsize\itshape,black!55] at (2.3,3.66) {wiggly, overlapping responses};
  \draw[poolgrey,dashed,line width=1pt] (0.35,2.05) -- (4.45,1.45);
  % regime A: concave saturating
  \draw[regA,line width=1.5pt] plot[smooth,tension=0.7] coordinates
     {(0.5,2.95)(1.2,2.70)(2.0,2.45)(2.9,2.30)(3.8,2.23)(4.4,2.21)};
  \foreach \x/\y in {0.6/2.90,1.3/2.62,2.1/2.50,3.0/2.22,4.1/2.26}
     \fill[regA] (\x,\y) circle (2.3pt);
  % regime B: large S-shape
  \draw[regB,line width=1.5pt] plot[smooth,tension=0.7] coordinates
     {(0.5,2.45)(1.2,1.70)(2.1,0.95)(3.0,1.50)(3.9,2.05)(4.4,2.15)};
  \foreach \x/\y in {0.7/2.30,1.5/1.50,2.2/1.00,3.1/1.60,4.0/2.00}
     \fill[regB] (\x,\y) circle (2.3pt);
  % regime C: wiggly, multiple fixed points
  \draw[regC,line width=1.5pt] plot[smooth,tension=0.7] coordinates
     {(0.5,1.35)(1.2,0.85)(2.0,1.42)(2.8,0.88)(3.7,1.38)(4.4,1.05)};
  \foreach \x/\y in {0.6/1.30,1.4/0.90,2.1/1.35,3.0/0.95,4.1/1.20}
     \fill[regC] (\x,\y) circle (2.3pt);
\end{scope}

%% ================= PANEL (iii): attractors diverge =================
\begin{scope}[xshift=11.8cm]
  \node[font=\small\bfseries] at (2.3,4.05) {(c) Attractors diverge};
  \node[font=\scriptsize\itshape,black!55] at (2.3,3.66) {separate basins of attraction};
  % three basins of attraction at divergent income positions
  \foreach \cx/\col/\lab in {0.85/regA/i,2.40/regB/ii,3.95/regC/iii}{
     \draw[\col,line width=1.5pt] plot[smooth,tension=0.7] coordinates
        {(\cx-0.70,3.00)(\cx-0.40,2.20)(\cx-0.15,1.70)(\cx,1.55)
         (\cx+0.15,1.70)(\cx+0.40,2.20)(\cx+0.70,3.00)};
     \draw[\col,line width=0.9pt,->] (\cx-0.42,2.35) -- (\cx-0.15,1.80);
     \draw[\col,line width=0.9pt,->] (\cx+0.42,2.35) -- (\cx+0.15,1.80);
     \fill[\col] (\cx,1.55) circle (4pt);
     \node[font=\scriptsize,black!65] at (\cx,3.28) {\lab};
  }
  % divergence span
  \draw[black!70] (0.85,1.30) -- (0.85,1.15) -- (3.95,1.15) -- (3.95,1.30);
  \node[font=\scriptsize,black!60] at (0.85,0.83) {$\sim\$100$};
  \node[font=\scriptsize,black!60] at (3.95,0.83) {$>\$30$k};
  \node[font=\scriptsize,black!60] at (2.40,0.30) {initial income (log scale)};
\end{scope}

\end{tikzpicture}
\caption{\textbf{Three objects conflated under ``convergence''.} \textbf{(i)}~Pooled $\beta$-convergence over all FUAs yields one slope. \textbf{(ii)}~By regime: regimes overlap in income but each converges to its own attractor. \textbf{(iii)}~Equal initial income, different regimes, divergent attractors spanning two orders of magnitude. The pooled slope just aggregates (ii) and (iii); divergence is in relative standing, every regime still grows.}
\label{fig:convergence_objects}
\end{figure}

The structure we recover is the one predicted by models with threshold externalities or non-convex production, in which initial conditions select among multiple basins of attraction \cite{azariadis1990threshold, durlauf1995multiple, Galor1996}; it extends the convergence-club evidence from regional data \cite{Quah1997, bartkowska2012regional} to a global panel of urban economies. It departs from that literature in one respect. Threshold models are usually read as models of poverty traps---low-level equilibria of absolute stagnation. We find no such trap at the urban scale: every regime grows in absolute terms. The divergence in Figure~\ref{fig:convergence_objects} is in \emph{relative} standing---the income at which an FUA's growth settles to its regime mean---not in absolute income. The regimes are evidence of persistent structural stratification, not of arrested growth, a distinction that separates them from the absolute-stagnation narrative of the middle-income trap \cite{eichengreen2012fast}.

\subsection{Implications for macro-stabilization}

Stabilization policy is built around the nation: one central bank, one macro-prudential authority, one fiscal rule, one seat at the regional facility. Growth regimes are not national. They are meso-scale, they cross borders, and they place FUAs of the same country in different positions of the propagation network. What this creates is not imprecision but a mismatch of units, with three consequences a stabilization authority cannot assume away. If regimes are a relevant operative unit of growth dynamics, and regimes cross national borders, then macroeconomic policy organized around the nation can be suboptimal in three ways. First, aggregation: in the third of countries whose FUAs span multiple regimes, a single national macro-prudential or industrial stance is calibrated to an average city that does not exist, and is necessarily wrong for FUAs embedded in different propagation roles \cite{barca2012case, RodriguezPose2022}. Second, geography: shock exposure follows structural similarity, so contagion policy that presumes spatial spread---regional bailouts, neighborhood firewalls---may firewall the wrong perimeter and miss a structurally linked regime on another continent \cite{Acemoglu2012, Starnini2019}. Third, buffering: buffer regimes contracted less than exporters or amplifiers in 2008--2009, but insulation is symmetric---the low integration that shields a regime from imported crises also excludes it from synchronized recoveries and mutually beneficial growth spurts. \\

These roles map onto familiar macroeconomic narratives, but only when read carefully. C3's net-exporter position is consistent with OECD demand and financial conditions setting the global cycle; C1's absorber position is consistent with a Chinese growth model driven, over this period, by external demand. The mapping is to the \emph{timing} of fluctuations---which regime moves first---not to trade balances: China was a shock absorber in our sense while running large current-account surpluses, and the two statements concern different objects. The ordering is also period-specific. 1993--2019 is one particular sequence of global conditions---the China boom, the post-Soviet transition, the global financial crisis, the commodity supercycle---and a different window could reorder the roles.

\subsection{Limitations}

Three limits bound the inference. The propagation network rests on lagged correlations, which establish temporal precedence but not causal channels; identifying mechanisms would require instrumental-variable or quasi-experimental designs we do not have. The regimes are recovered from cross-sectionally standardized growth and therefore describe trajectories of \emph{relative} rank; an across-time standardization that clusters on trajectory shape yields a substantially different partition (ARI~$=0.39$), and the regimes should be read as relative-rank objects accordingly. And the clustering is uneven in robustness: claims are strongest for the anchor regimes and weakest where no stable cluster occupies a role---the amplifier role, in particular, is not held by a reproducibly identified regime.\\

Two further caveats concern data and scope. Nighttime-light GDP is an imperfect proxy \cite{bickenbach2016night, mellander2015night}, although regime assignments are 94\% stable on the higher-income subsample where calibration is most reliable, and the 28-fold spread in regime volatility exceeds any plausible measurement-noise bound. Mature economies fall almost entirely within a single regime; this may reflect real homogeneity in mature urban systems, convergence already achieved, or saturation of the proxy in fully electrified countries, and we cannot adjudicate among the three. We also treat regimes as 1993--2019 attractors and do not test whether individual FUAs transitioned between them---a question the present data could address, and the natural next step. Finally, the ecological-niche metaphor \cite{Hidalgo2007} is licensed only in part: the regimes are real, persistent, and differ in behavior and network influence, but we do not identify the competition, displacement, or capability dynamics that a fully ecological reading would require.\\

The mesoscale lens refines the national one; it does not replace it. Clustering on growth trajectories alone, with no geographic or political input, recovers country and regional blocks leaving roughly two-thirds of countries internally coherent. This is evidence that the nation is a legible unit of analysis: for most of the world, national institutions organize the growth of cities. The growth behaviour of the global FUA system provides a lens that bridges both the supra and the subnational scale.

\section{Methods}
We combine two primary data sources: (i) globally gridded GDP estimates at 30 arc-second resolution (\textasciitilde1 km at the equator) covering 1993--2019 \cite{Chen2022}, and (ii) global FUA boundaries from the 2015 Copernicus Global Human Settlement Layer Functional Urban Areas (GHS-FUA) dataset \cite{Florczyk2019}. The gridded GDP data are derived from nighttime light (NTL) intensity measurements calibrated against national accounts and expressed in constant 2011 PPP-adjusted USD. We use the data at their lowest spatial resolution and full temporal coverage to ensure a homogeneous panel with minimal interpolation artifacts. GDP values are used directly without further deflation or scaling to preserve the magnitude of cross-sectional differences. FUA boundaries are taken from the 2015 layer and treated as time-invariant throughout the analysis period. This choice reflects our interest in the long-run growth dynamics of stable urban systems rather than short-run boundary changes. Each FUA is defined as a city plus its commuting zone, based on a 15\% commuting threshold to the urban center \cite{Dijkstra2019}, capturing the functional economic extent of urban agglomerations more accurately than administrative boundaries.

\subsection{Spatial aggregation procedure}

To align the gridded GDP raster with FUA polygons, we first reproject both datasets to a common equal-area coordinate reference system (CRS), specifically World Mollweide projection (EPSG:54009). This ensures that area-based calculations are consistent across latitudes and that the spatial aggregation does not introduce systematic distortions.

For each FUA $i$ and year $t$, we construct aggregate GDP as the sum of all raster cells whose centroids lie within the FUA polygon:
\begin{equation}
\text{GDP}_{i,t} = \sum_{j \in \mathcal{C}_i} \text{GDP}_{j,t},
\end{equation}
where $\mathcal{C}_i$ denotes the set of all grid cells $j$ whose centers fall inside FUA $i$. Cells intersecting the boundary are fully assigned to the FUA if their center falls inside; otherwise they are excluded. Given the coarse resolution of the grid ($\sim$1 km) relative to typical FUA sizes (tens to hundreds of square kilometers) and the rarity of boundary-sharing between neighboring FUAs in the global dataset, the resulting over- or under-coverage error is negligible compared to the conceptual uncertainty inherent in urban boundary definitions themselves.

We exclude FUAs with no valid GDP data in any year over the 1993--2019 period, yielding a final panel of $N = 8{,}808$ FUAs across $C = 165$ countries and $T = 27$ years.

\subsection{Growth trajectory construction and regime clustering}

Annual growth rates are computed as discrete percentage changes in GDP:
\begin{equation}
g_{i,t} = \frac{\text{GDP}_{i,t} - \text{GDP}_{i,t-1}}{\text{GDP}_{i,t-1}} \times 100,
\end{equation}
for $t = 1994, \ldots, 2019$. This formulation preserves the asymmetry between contractions and booms (e.g., a $-50\%$ shock and a $+50\%$ shock are not treated as symmetric), which is economically meaningful. We do not smooth, detrend, or filter these series. Retaining the raw volatility is essential because regimes are defined by their full temporal ``shape'', including crises, recoveries, and structural breaks. Any pre-processing that removes volatility would obscure the very patterns we seek to identify.

Each FUA $i$ is then represented by its growth trajectory vector:
\begin{equation}
\mathbf{v}_i = (g_{i,1994}, g_{i,1995}, \ldots, g_{i,2019}) \in \mathbb{R}^{26}.
\end{equation}
This vector captures the entire temporal signature of FUA $i$'s economic dynamics over the sample period.

\subsection{Preprocessing and dimensionality reduction}
Two preprocessing steps precede clustering. First, growth rates are standardised \emph{cross-sectionally}: for each year $t$, each FUA's growth rate $g_{i,t}$ is expressed as a $z$-score relative to the distribution $\{g_{j,t}\}_{j=1}^N$ across all FUAs in that year. This removes year-specific mean and scale effects (in particular the common-mode component of any global shock that moves all FUAs in the same direction), so that what remains is each FUA's position relative to its contemporaries. Regimes are then recovered from these cross-sectional ranks rather than from absolute growth levels.

Second, we reduce the dimensionality of the standardised growth matrix  $\tilde{\mathbf{V}} \in \mathbb{R}^{N \times 26}$  via principal component analysis (PCA), retaining the first $d$ components that jointly explain at least 80\% of the total variance. In our sample this yields $d = 14$ components. The retained principal-component scores $\mathbf{w}_i \in \mathbb{R}^{14}$ preserve boom--bust cycles, trend breaks, and volatility clustering while discarding noise-dominated dimensions. Because PCA is a linear, invertible transformation (up to the discarded components), it does not distort the relative geometry of trajectories in the retained subspace.

\subsection{Clustering via k-means}
We partition the PCA-reduced representations $\{\mathbf{w}_i\}_{i=1}^N$ into $k$ clusters using k-means \cite{Lloyd1982} with k-means++ initialisation \cite{Arthur2007}, as implemented in scikit-learn \cite{Pedregosa2011}. During cluster-number search we use 10 restarts per $k$; for the final fit at $k^*$, we use 20 restarts to improve solution quality.

\subsection{Optimal cluster selection via silhouette score}
The number of clusters $k$ is selected empirically by maximising the mean silhouette coefficient $\bar{s}(k)$ \citep{Rousseeuw1987}, which jointly measures cluster cohesion and separation in the PCA-reduced space in which clustering is performed. 

We sweep $k = 3, \ldots, 20$ with 20 independent random seeds per $k$ and select $k^* = \operatorname{argmax}_k \bar{s}(k)$, where $\bar{s}(k)$ is averaged across seeds (Supplementary Figure ~\ref{fig:Silhoutte}). This procedure yields $k^* = 17$ with mean silhouette $0.267 \pm 0.013$. The final cluster assignment is obtained by re-fitting k-means at $k^* = 17$ using the seed that maximised silhouette during the sweep; all downstream analyses (regime trajectory characterisation, shock detection, propagation network construction) are based on these 17 clusters.

% We evaluate the mean silhouette $\bar{s}(k)$ for $k = 3, \ldots, 20$ . To account for sensitivity to initialisation, each $k$ is evaluated across 20 independent random seeds, and we report the mean and standard deviation of the silhouette score across seeds. The optimal number of clusters is selected as $k^* = \operatorname{argmax}_{k} \bar{s}(k)$, where $\bar{s}(k)$ is averaged over all seeds. This procedure yields $k^* = 17$ clusters with mean silhouette $0.267 \pm 0.013$.

The absolute silhouette coefficient is moderate, as expected, given the continuous nature of growth dynamics: urban economies do not fall into sharply separated types but rather occupy overlapping regions of trajectory space, with regime boundaries representing zones of gradual transition rather than hard discontinuities. The silhouette score should therefore be evaluated relative to a null benchmark rather than against an idealized threshold. To verify that these clusters reflect genuine temporal structure rather than artefacts of high-dimensional noise, we compare the observed silhouette scores against a null distribution constructed by independently permuting the year ordering within each FUA's growth trajectory (50 permutations), re-standardising and re-projecting via PCA, and re-clustering. The null model yields a mean silhouette of $0.053 \pm 0.005$ at $k = 17$---a fivefold difference---confirming that the identified regimes capture meaningful temporal coherence far exceeding what random trajectories produce. Further robustness to alternative clustering approaches is assessed in Supplementary~\ref{sec:permutation_null}. For each cluster $c = 0, 1, \ldots, 16$, we compute the mean growth trajectory as the simple average across all member FUAs:
\begin{equation}
\bar{g}_{c,t} = \frac{1}{|\mathcal{S}_c|} \sum_{i \in \mathcal{S}_c} g_{i,t},
\end{equation}
where $|\mathcal{S}_c|$ is the number of FUAs in cluster $c$. These cluster-level trajectories serve as the primary objects for characterising regime behaviour and identifying extreme events.

\subsection{Cluster stability and core-set construction}
\label{sec:stability}

An FUA enters the \emph{core set} if its k-means regime assignment is corroborated by at least one of two independent clustering methods: HDBSCAN \citep{Campello2013} or a Gaussian Mixture Model (GMM). Per-regime Jaccard is the agreement statistic, computed after Hungarian matching of cluster labels to the k-means reference partition \citep{Munkres1957}; equivalently, an FUA is excluded only if \emph{both} alternatives disagree with k-means. This rule yields a core set of $N_{\text{core}} = 7{,}474$ FUAs ($84.8\%$ of the sample), spanning all 17 regimes; a stricter rule requiring unanimous agreement retains $4{,}099$ FUAs ($46.5\%$). All regime-level analyses reported in the main text are computed on the core set. Hyperparameter settings for HDBSCAN and GMM, the treatment of HDBSCAN's noise-flagged FUAs, and the exclusion of spectral clustering from the core-set construction are detailed in Supplementary~\ref{sec:stability_apparatus}.

Per-regime Jaccard indices (Supplementary Table~\ref{tab:jaccard}) identify four regimes whose membership is recovered consistently at the chosen $k = k^* = 17$: C1 (China-dominated), C4 (Venezuela), C7 (Iraq), and C13 (Iran). This is a fixed-$k$ criterion; under joint perturbation of both method \emph{and} $k$, C13 merges into a larger cluster at $k = 18$, while C1, C4, and C7 remain stable. A label-free \emph{super-core} measure that formalises joint-perturbation stability is reported in Supplementary~\ref{sec:supercore}, and a complementary bootstrap-resample stability test \citep{Hennig2007} on $B = 100$ FUA-bootstrap samples is reported in Supplementary~\ref{sec:bootstrap_jaccard}.

\subsection{Shock detection and propagation network}
Shocks are defined as the most extreme local fluctuations in each regime's mean growth series, using a regime-specific threshold that respects volatility heterogeneity: what is exceptional in a low-variance mature regime may be routine in a high-variance structurally unstable one. For each cluster $c$ we compute the long-run mean $\mu_c$ of its annual growth series $\bar{g}_{c,t}$ (with $T = 26$ growth-rate years over 1994--2019, since growth rates require a one-year lag from the 27-year GDP panel) and the deviations $\delta_{c,t} = \bar{g}_{c,t} - \mu_c$. Year $t$ is flagged as a shock for regime $c$ if $\delta_{c,t}$ falls below the 2nd or above the 98th percentile of the regime's own deviation distribution; negative shocks correspond to severe contractions, positive shocks to booms. Regime volatility $\sigma_c$ is computed as the sample standard deviation of $\bar{g}_{c,t}$ and is used to interpret the propagation roles described below.

\subsection{Lagged correlation and propagation network construction}
To examine propagation across regimes, we compute Pearson cross-correlations $\rho_{cc'}(\tau)$ between $\bar{g}_{c,t}$ and $\bar{g}_{c',t+\tau}$ for every ordered pair $(c, c')$ and lag $\tau \in \{0, 1, 2, 3\}$ years. Significance uses the standard large-sample approximation with $\text{SE}(\rho) \approx 1/\sqrt{n}$ under $H_0: \rho = 0$ and $n = T - \tau$ effective observations; we retain only edges with $|\rho_{cc'}(\tau)| > 1.96/\sqrt{n}$, corresponding to $|\rho| > 0.40$ at the longest lag ($n = 23$).

We do not correct the per-edge threshold for multiple comparisons; instead, the validity of the network ensemble is established by a 10{,}000-permutation null on the regime-level growth series (observed 21 inter-regime edges versus expected null $2.7 \pm 1.6$, $p < 0.001$; Section~\ref{sec:permutation_null}). The per-edge threshold corresponds to $p \approx 0.05$ at $n = 23$, applied across $12 \times 11 \times 4 = 528$ ordered (regime, lag) cells; a per-edge multiple-testing audit on the parallel 17-regime correlation matrix is reported in SI figure ~\ref{fig:growth_heatmap_full}.

The set of statistically significant directed correlations forms a weighted, directed propagation network $\mathcal{G} = (\mathcal{V}, \mathcal{E})$, where $\mathcal{V}$ is the set of 17 regimes and an edge $(c \to c', \tau)$ with weight $\rho_{cc'}(\tau)$ indicates that growth fluctuations in regime $c$ at time $t$ are predictive of fluctuations in regime $c'$ at time $t + \tau$.

\subsection{Propagation roles and network metrics}
For each regime $c$ we summarise its position in $\mathcal{G}$ by two scalar statistics: $\text{Out}_c$, the sum of all significant outgoing edge weights $\rho_{cc'}(\tau)$ across $c' \neq c$ and lags $\tau \in \{0, 1, 2, 3\}$, and $\text{In}_c$, the analogous sum over significant incoming edges. Net export strength is then $\text{NetExport}_c = \text{Out}_c - \text{In}_c$. Each regime is classified into one of four roles by whether its $\text{Out}_c$ and $\text{In}_c$ lie above or below the cross-regime medians: \textbf{Exporters} (high out, low in), \textbf{Absorbers} (low out, high in), \textbf{Amplifiers} (high both), and \textbf{Buffers} (low both). Exporters display growth fluctuations that temporally precede those of other regimes; absorbers display fluctuations that temporally follow them; amplifiers exhibit strong co-movement in both directions; and buffers show weak co-movement overall. These labels describe patterns of temporal precedence, not verified causal channels.

\section*{Data availability}

The gridded GDP dataset is available at \cite{Wang2022}. FUA boundaries are available from \cite{ghs_fua_2023}. Derived regime assignments are available from the corresponding author upon reasonable request.

\section*{Code availability}

The code used for data processing, clustering, and network construction is available at GitHub. An archived version corresponding to this manuscript will be deposited on Zenodo prior to publication.
\href{https://github.com/Tarsatir/mesoscales-of-growth}{https://github.com/Tarsatir/mesoscales-of-growth}

\section*{Acknowledgements}

The authors acknowledge the support from the Dutch Research Council (NWO). Roy, D. (2024). Leveraging nighttime light data to reveal global economic dynamics [https://doi.org/10.61686/TOTZO12323]. Dutch Research Council (NWO). .

\section*{Author contributions}

I.M. conceived the study, designed and implemented the analysis, and performed the data processing. I.M. and D.R. jointly developed the theoretical framing, interpreted the results, and wrote the manuscript. C.H. designed the robustness analyses and revised the Methods and Supplementary Information. All authors revised the manuscript. D.R. supervised the research.

\section*{Competing interests}

The authors declare no competing interests.

\bibliographystyle{plainnat}
\bibliography{sample}
\newpage

\section{Supplementary Information}
\FloatBarrier

\subsection{Cluster growth regimes}
For each regime we construct year-$t$ vs.\ year-$(t+1)$ return maps of per-capita GDP, with a LOWESS smooth of the conditional expectation $\mathbb{E}[y_{t+1}\mid y_t]$ and the trend diagonal $y = x + \bar{g}_c$. Intersections of the smooth with the diagonal are fixed points: within-regime equilibria where conditional growth equals the regime mean. Deviations of the smooth from the diagonal are magnified $\times 30$ for visibility.

Regimes differ in the shape of $\mathbb{E}[y_{t+1}\mid y_t]$, not only in the mean trajectory $\bar{g}_{c,t}$ characterised in Section 2.2. C1 (China), C3 (OECD+), and C13 (Iran/Syria) show concave bending above $\sim$\$5K, consistent with within-regime saturation. C1 and C13 are cross-method-stable (Methods §4.5), and C3 is stable under bootstrap and super-core resampling (Sections 5.8, 5.10), so the saturation signal concentrates in the better-corroborated regimes. C14 (Southeast Asia) shows a single-fixed-point S-shape with curvature both below and above its fixed point, and C0 (Dynamic Emerging) shows weaker concavity at the upper end. Several regimes show multiple fixed points: C9 (Nigeria+), C16 (Ethiopia), and the low-stability regimes C10 (Steady Emerging) and C15 (DRC+Zimbabwe), which fail the bootstrap and super-core tests of Sections 5.8 and 5.10. Multiple crossings reflect either multi-modal within-regime income distributions or stretches where the smooth tracks the diagonal closely, and should be read cautiously given the $\times 30$ magnification. C15 is the only regime with a sigmoid-like double bend, but with $n=138$ conflict-dominated FUAs its shape is not robustly identified. Fixed-point locations span roughly two orders of magnitude, from $\sim$\$100 in C9 to $>$\$30K in C3, indicating that regimes occupy distinct income-level attractors rather than tracing a single global development path.

These conditional dynamics are descriptive. They locate where within-regime drift vanishes but do not identify the structural primitives (savings rates, capital intensity, productive capabilities, integration into global value chains) that set fixed-point location; separating candidate mechanisms requires subnational data on these primitives, not currently available at FUA resolution at global scale. Three theoretical traditions are consistent with the observed pattern: Solow-type models with regime-specific parameters \citep{Solow1956, MankiwRomerWeil1992}, in which savings rates and effective depreciation shift the steady state; path-dependence and lock-in models \citep{Arthur1989, David1985, Dosi1982}, in which historical contingency selects among basins of attraction; and capability-based development \citep{Hidalgo2007}, in which productive capability sets cap regimes at distinct income ceilings. The data are consistent with all three, and distinguishing among them is left to future work.

\begin{figure}[h]
    \centering
    \includegraphics[width=0.95\linewidth]{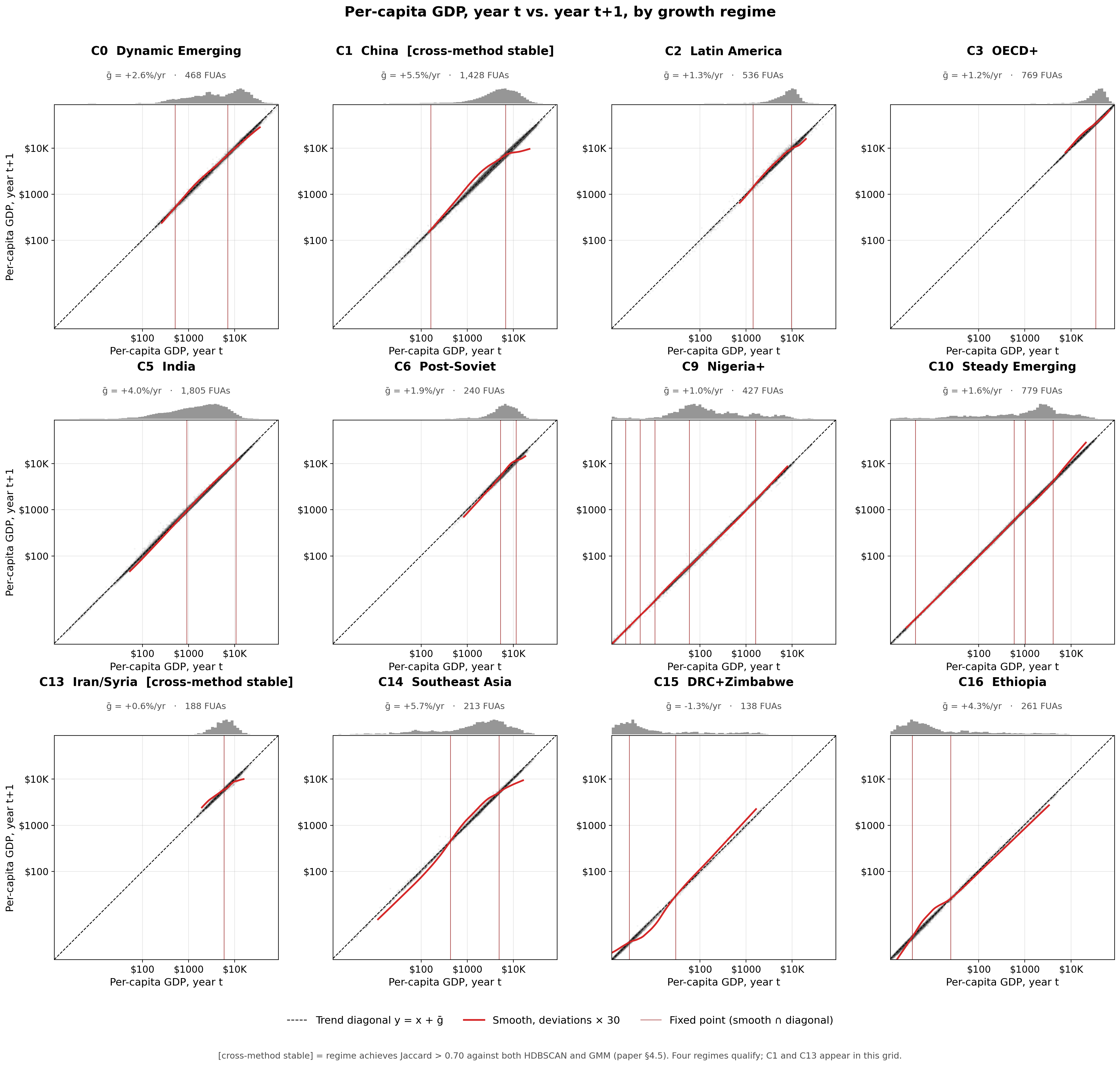}
    \caption{Cluster exhibit different behavior in growth regimes across GDP levels. Varying in fixpoint location and number of fixpoints. }
    \label{fig:cluster_regimes}
\end{figure}

\FloatBarrier
\subsection{Industrialization timing and within-country spatial inequality}

\begin{figure}[h]
    \centering
    \includegraphics[width=0.75\textwidth]{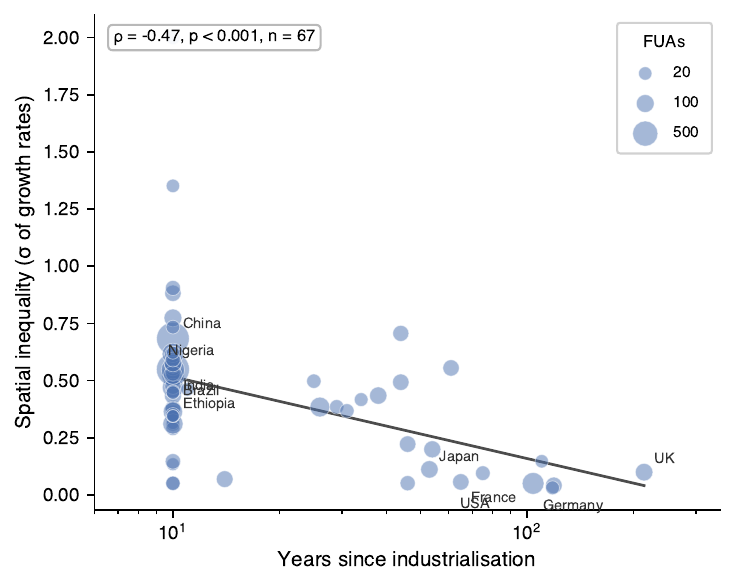}
    \caption{\textbf{Within-country spatial inequality in FUA growth rates declines
    with industrialization maturity.} Each point represents a country with multiple
    FUAs; point size is proportional to the number of FUAs. The $x$-axis shows years
    since the country first had industrial employment exceed agricultural employment
    (log scale). The $y$-axis shows the cross-sectional standard deviation of FUA
    growth rates, averaged over 1993--2019.
    The fitted line follows a log-linear specification.}
    \label{fig:indust_inequality}
\end{figure}

\begin{figure}
    \centering
    \includegraphics[width=1\linewidth]{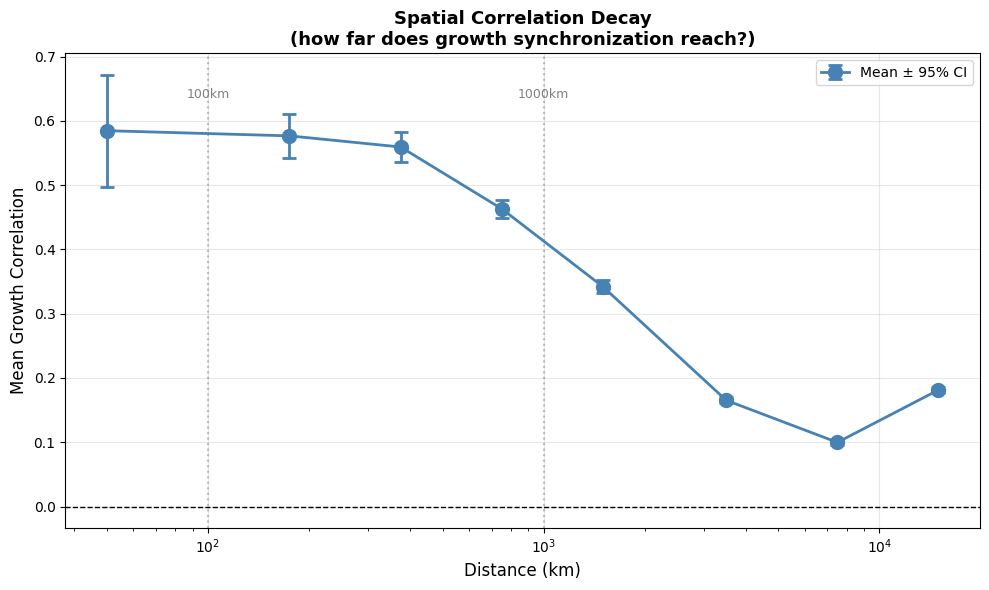}
    \caption{
        \textbf{Propagation strength decays weakly with geographic distance.}
        Maximum absolute lagged cross-cluster correlation ($\tau = 0, \ldots, 3$ years) versus great-circle distance between cluster centroids (thousands of km) for all 136 regime pairs. The weak negative slope and substantial scatter indicate that geographic distance is a poor predictor of propagation strength: many distant pairs ($\geq$10,000~km) exhibit correlations $\geq$0.4, while some proximate pairs show near-zero correlation. Structural similarity in production composition and financial integration explains the residual variance more effectively than geographic proximity.
    }
    \label{fig:spatialdecay}
\end{figure}

\begin{figure}
    \centering
\includegraphics[width=1\linewidth]{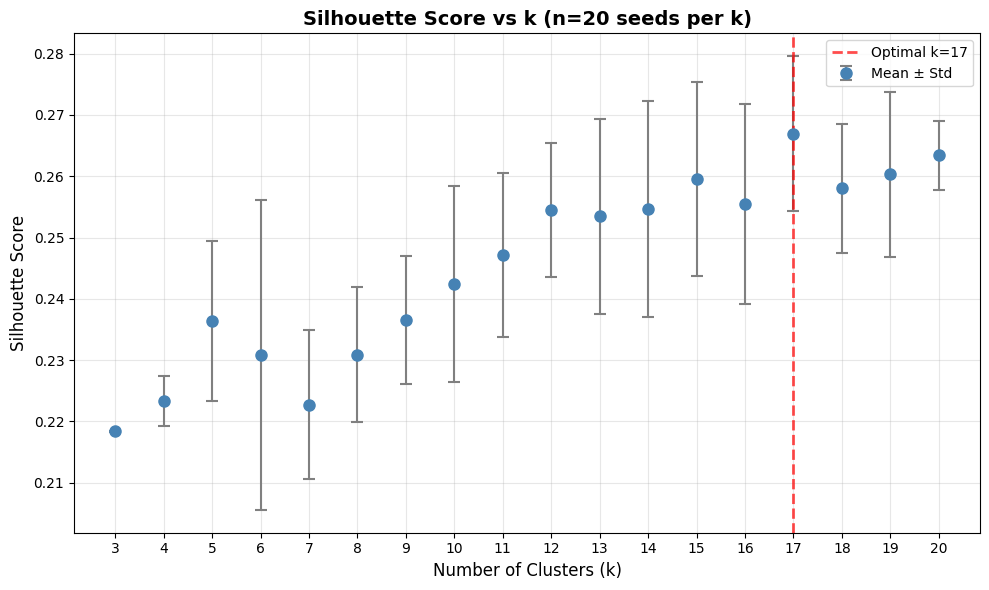}
        \caption{Silhouette analysis across candidate numbers of clusters 
$k \in \{3,\dots,20\}$ using $20$ random initialisations per $k$. 
Dots report the mean silhouette coefficient $\bar{s}(k)$ and error bars 
its standard deviation. The function $\bar{s}(k)$ attains its maximum at 
$k = 17$ (red dashed line), which we therefore select as the optimal 
number of clusters.
    }
    \label{fig:Silhoutte}
\end{figure}

If the growth regimes identified in the Results capture meaningful economic structure, we would expect their spatial distribution within countries to reflect deeper structural features of national development. We test this by examining the relationship between industrialization timing and within-country spatial inequality in growth rates. For each country with multiple FUAs, we compute the cross-sectional standard deviation of FUA-level GDP growth rates, averaged over the sample period, as a measure of within-country spatial inequality. We relate this to the number of years since each country first had industrial employment exceed agricultural employment~\cite{bentzen2013timing}, restricting the sample to the 67 countries for which both industrialization-timing data and multiple FUAs are available.Figure~\ref{fig:indust_inequality} shows a significant negative relationship (Spearman $\rho = -0.47$, $p < 0.001$, $n = 67$ countries): countries at earlier stages of industrialization exhibit markedly greater within-country spatial inequality in FUA growth rates. Recently industrializing countries display dispersed growth patterns---with different cities following distinct growth trajectories---while mature industrial economies show more spatially homogeneous growth. This is consistent with the hypothesis that early-stage development concentrates growth in a few leading cities before spreading more evenly across the urban system.

\FloatBarrier

% Table 1: Cluster Descriptive Statistics
\begin{table}[htbp]
\centering
\caption{Cluster Descriptive Statistics and Shock Propagation Roles}
\label{tab:cluster_stats}
\small
\begin{tabular}{crrrrl}
\toprule
\textbf{Cluster} & \textbf{N FUAs} & \textbf{N Countries} & \textbf{Net Export} & \textbf{Volatility} & \textbf{Role} \\
\midrule
0 & 589 & 16 & 0.018 & 2.78 & Amplifier \\
1 & 1,430 & 7 & -0.066 & 2.25 & Absorber \\
2 & 627 & 20 & -0.007 & 3.01 & Amplifier \\
3 & 1,208 & 42 & 0.051 & 1.45 & Exporter \\
4 & 64 & 1 & -0.018 & 5.64 & Amplifier \\
5 & 1,880 & 5 & 0.028 & 1.71 & Buffer \\
6 & 322 & 9 & 0.009 & 4.82 & Amplifier \\
7 & 64 & 1 & -0.014 & 18.93 & Amplifier \\
9 & 449 & 6 & -0.006 & 2.07 & Buffer \\
10 & 1,015 & 42 & 0.031 & 0.66 & Buffer \\
11 & 129 & 3 & -0.056 & 1.80 & Absorber \\
12 & 24 & 3 & 0.009 & 7.91 & Amplifier \\
13 & 188 & 2 & -0.010 & 3.79 & Amplifier \\
14 & 402 & 5 & -0.015 & 1.34 & Buffer \\
15 & 142 & 2 & 0.039 & 2.18 & Buffer \\
16 & 275 & 3 & 0.005 & 3.40 & Amplifier \\
\midrule
\textbf{Total} & \textbf{8,808} & \textbf{165} & -- & -- & -- \\
\bottomrule
\end{tabular}
\vspace{0.5em}

\noindent\textit{Note:} Cluster~8 is absent because k-means labels are assigned based on cluster size (largest = 0), and the original 18-cluster solution merged two nearly identical clusters during final refinement, leaving the label sequence non-contiguous. The 17 reported clusters represent the final, stable partition.
\end{table}

% \textit{Notes:} Shocks detected using bottom/top 2\% threshold within each cluster's growth trajectory (1993-2019). Net Export = outgoing shock strength minus incoming shock strength. Volatility = standard deviation of cluster growth rate. Roles assigned based on quadrant position: Exporters (high outgoing, low incoming), Absorbers (low outgoing, high incoming), Amplifiers (high both), Buffers (low both).

% Table 2: Country Allocation to Clusters
\begin{table}[htbp]
\centering
\caption{Country Allocation to Growth Clusters}
\label{tab:country_allocation}
\small
\begin{tabular}{cp{12cm}}
\toprule
\textbf{Cluster} & \textbf{Countries} \\
\midrule
0 & Indonesia, Turkey, Philippines, Thailand, Malaysia, Peru, South Korea, Botswana, Sri Lanka, Hong Kong, Malawi, UAE, Djibouti, Mongolia, Montenegro, Singapore \\
\addlinespace
1 & China, Angola, Armenia, Macao, North Korea, Qatar, Laos \\
\addlinespace
2 & Brazil, Colombia, Argentina, South Africa, Ecuador, Chile, Yemen, Oman, Lebanon, Paraguay, Uruguay, Trinidad \& Tobago, Palestine, Kuwait, Lesotho, Comoros, Cape Verde, Suriname, Swaziland, Namibia \\
\addlinespace
3 & United States, Mexico, United Kingdom, Japan, Italy, Germany, France, Spain, Algeria, Canada, Saudi Arabia, Netherlands, Haiti, Switzerland, Burundi, Sweden, Greece, Belgium, Hungary, Nicaragua, El Salvador, Portugal, Austria, Finland, Kyrgyzstan, Madagascar, CAR, Croatia, Israel, Macedonia, Jamaica, Norway, Denmark, Cyprus, Guinea-Bissau, Gabon, Slovenia, Luxembourg, Brunei, Barbados, Bahamas, Belize \\
\addlinespace
4 & Venezuela \\
\addlinespace
5 & India, Burkina Faso, Uganda, Albania, Ireland \\
\addlinespace
6 & Russia, Ukraine, Kazakhstan, Azerbaijan, Belarus, Tajikistan, Lithuania, Moldova, Latvia \\
\addlinespace
7 & Iraq \\
\addlinespace
9 & Nigeria, Chad, Serbia, Sierra Leone, Republic of Congo, São Tomé \& Príncipe \\
\addlinespace
10 & Pakistan, Egypt, Morocco, Bangladesh, Uzbekistan, Cameroon, Poland, Sudan, Guatemala, Romania, Tanzania, Kenya, Niger, Senegal, Zambia, Ghana, Guinea, Australia, Czech Republic, Tunisia, Nepal, Honduras, Bolivia, Jordan, Dominican Republic, Bulgaria, New Zealand, Taiwan, Slovakia, Mauritania, Gambia, Costa Rica, Panama, Estonia, Georgia, Fiji, Mauritius, Malta, Bahrain, Iceland \\
\addlinespace
11 & Côte d'Ivoire, Benin, Togo \\
\addlinespace
12 & Bosnia \& Herzegovina, Liberia, Equatorial Guinea \\
\addlinespace
13 & Iran, Syria \\
\addlinespace
14 & Myanmar, Vietnam, Mozambique, Mali, Cambodia \\
\addlinespace
15 & DR Congo, Zimbabwe \\
\addlinespace
16 & Ethiopia, Turkmenistan, Rwanda \\
\bottomrule
\end{tabular}
\end{table}

\subsection{Permutation null for the propagation network}
\label{sec:permutation_null}
\begin{figure}[t]
    \centering
    \includegraphics[width=0.65\textwidth]{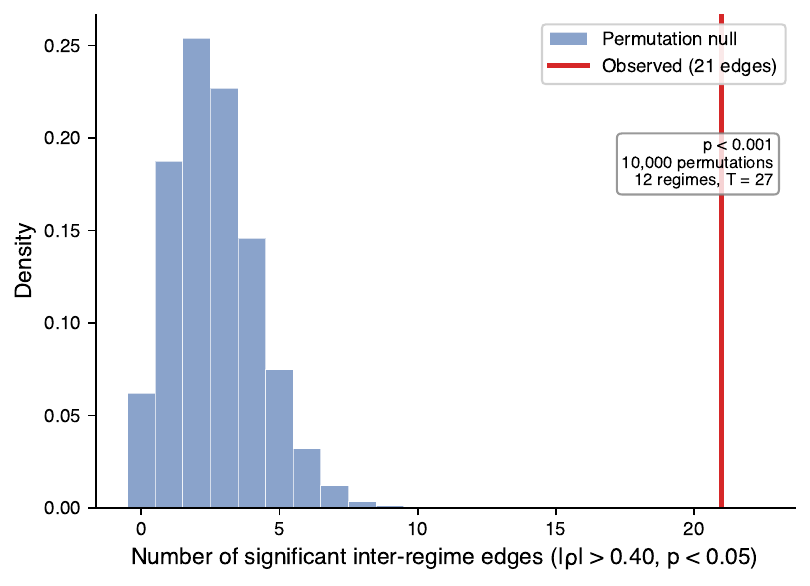}
    \caption{\textbf{The inter-regime propagation network is significantly
    non-random.} Distribution of significant edges ($|\rho| > 0.40$, $p < 0.05$)
    from 10,000 temporal permutations of regime-level growth time series.
    The red line marks the observed count (21 edges). Under the null,
    the mean is 2.7 edges.}
    \label{fig:network_null}
\end{figure}

To confirm that the inter-regime propagation network reflects genuine temporal dependencies rather than statistical artefacts, we compare the observed network to a permutation null. We compute regime-level mean growth time series by averaging across all FUAs in each regime, yielding 12 time series (after excluding five smaller regimes to ensure stable mean trajectories), each of length $T = 26$ annual growth-rate observations. The observed network among these 12 regimes contains 21 significant inter-regime edges ($|\rho| > 0.40$, $p < 0.05$) out of 66 possible directed pairs. We then independently permute the temporal ordering of each regime's growth series (10,000 permutations), destroying any lead--lag structure while preserving marginal distributions. Under this null, the expected number of significant edges is $2.7 \pm 1.6$ ($p < 0.001$; Figure~\ref{fig:network_null}). The observed propagation structure is thus overwhelmingly unlikely to arise by chance, confirming that the temporal correlations between growth regimes reflect real economic linkages.

\FloatBarrier
\subsection{Regime decomposition of Countries}
\begin{figure}[t]
    \centering
    \includegraphics[width=0.85\textwidth]{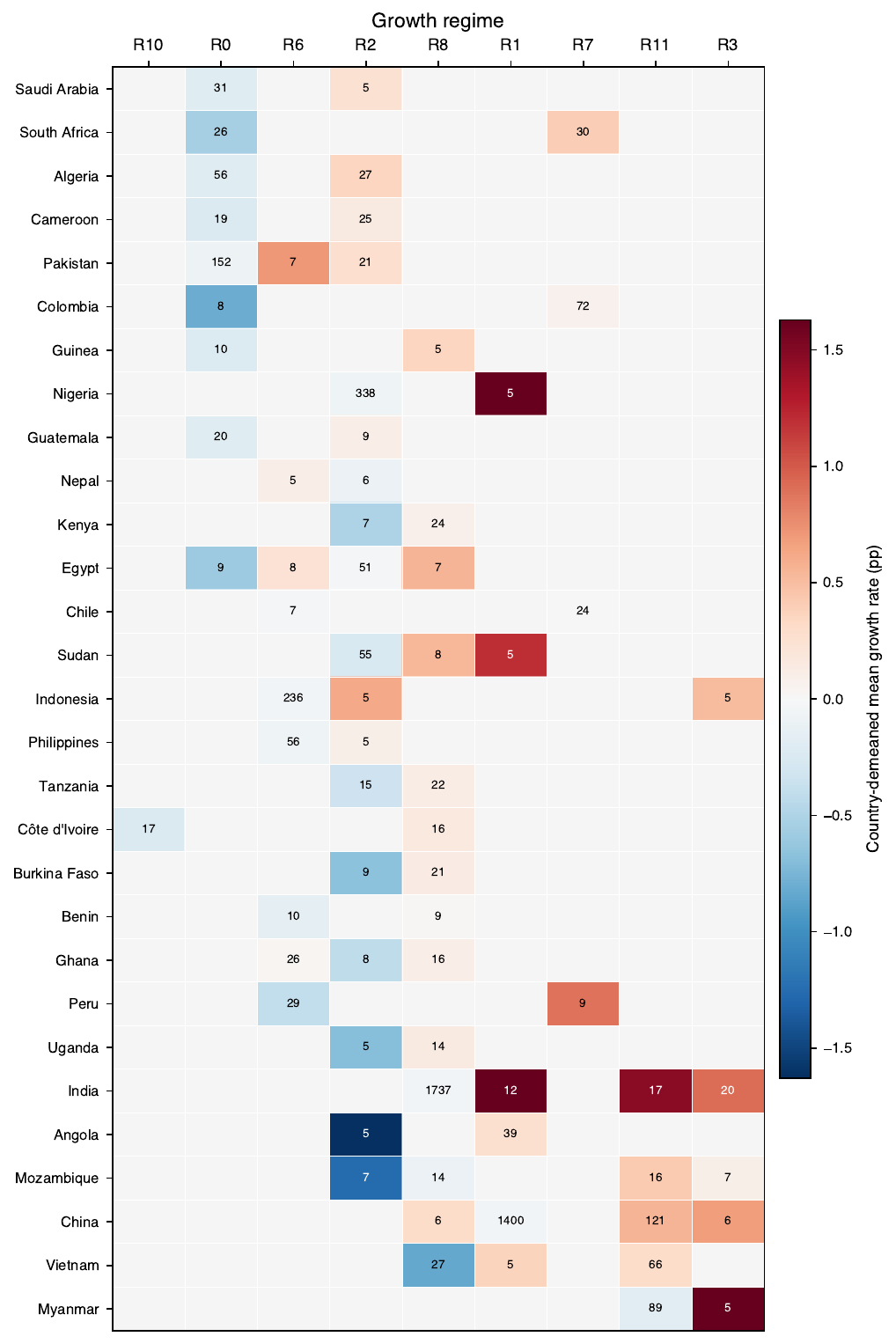}
    \caption{\textbf{Country-demeaned FUA growth rates by regime for all qualifying
    countries.} Showing all 29 countries (of 165 in the full sample) with at least two regimes each containing five or more FUAs.
    Countries are sorted by mean national growth rate (top = slowest).
    Cell numbers indicate FUA counts.}
    \label{fig:growth_heatmap_full}
\end{figure}

To test whether growth regime membership captures economically meaningful variation beyond national-level effects, we decompose the variance of FUA-level growth rates into contributions from country and regime membership. We restrict the analysis to the 85 countries whose FUAs span multiple regimes (6,736 of the full-sample 8,808 FUAs; single-regime countries are excluded because within-country regime variation is undefined for them). Adding regime membership to country fixed effects significantly improves the model ($F(11,6640) = 114.25$, $p < 0.001$, partial~$\eta^2 = 0.159$): regime membership explains 16\% of within-country variance in FUA-level mean growth rates. The result also holds in a full FUA$\times$year panel with country and year fixed effects ($F(11,181750) = 71.62$, $p < 0.001$). Figure~\ref{fig:growth_heatmap_full} illustrates this for the 29 countries (of 165 in the full sample) with at least two regimes each containing five or more FUAs. Within each country, FUAs assigned to different regimes show systematically different growth rates: regimes toward the right of the figure (R7, R11, R3) consistently grow faster than the country average, while those toward the left (R0, R6) grow slower. The pattern is consistent across countries spanning different continents and income levels. Notably, mature industrial economies (United States, Germany, Japan, United Kingdom) are absent because their FUAs fall almost entirely within a single regime---an observation consistent with the declining spatial inequality documented in Section~4.1. 

\FloatBarrier
\subsection{Robustness of the clustering pipeline}
\label{sec:robustness}

The baseline clustering pipeline (PCA to 80\% variance, k-means with silhouette selection) involves several methodological choices. We assess sensitivity along three dimensions.

\subsubsection{PCA variance threshold.} Varying the retained variance from 70\% ($d = 10$ components) to 90\% ($d = 18$) yields silhouette-optimal solutions of $k = 15$ to $k = 19$ clusters. The Adjusted Rand Index (ARI) between the baseline 17-cluster solution and these alternatives exceeds 0.82 in all cases, indicating that the broad regime structure is robust to the dimensionality of the PCA subspace.

\subsubsection{Alternative clustering algorithms.} Beyond the core-set assessment in Methods~\ref{sec:stability}, we additionally re-cluster the baseline PCA scores using hierarchical agglomerative clustering with Ward linkage \citep{Ward1963} and DBSCAN ($\varepsilon$ selected by the knee method on the $k$-distance graph, $\text{minPts}=15$). Ward-linkage clustering at $k=17$ achieves Adjusted Rand Index \citep{HubertArabie1985} $= 0.79$ against the k-means solution; DBSCAN identifies 14 core clusters (plus a noise class containing 6\% of FUAs) with $\mathrm{ARI} = 0.71$. Per-regime Jaccard indices for the core-set methods (HDBSCAN, GMM) are reported in Table~\ref{tab:jaccard}; the propagation network structure (exporter, absorber, amplifier, and buffer classifications) is qualitatively unchanged when recomputed on the core set alone.

\subsubsection{NTL measurement noise.} To assess whether NTL-specific measurement error drives the clustering, we restrict the sample to FUAs with initial GDP above the global median, where NTL calibration is most reliable. Re-running the full pipeline on this subsample ($N = 4{,}404$ FUAs) recovers 15 clusters with ARI = 0.87 against the original assignments of these FUAs, confirming that regime structure is not an artefact of low-income measurement noise.

\subsubsection{HDBSCAN noise as disagreement}
\label{sec:hdbscan_noise_audit}

Of the $N = 8{,}808$ FUAs in the sample, 3{,}754 (42.6\%) are labelled as noise by HDBSCAN under the canonical hyperparameters. Our reference core-set rule yields the 84.8\% corroborated-core fraction reported in Methods~\ref{sec:stability} by reassigning these noise labels to the nearest k-means centroid before Hungarian matching. A stricter rule, in which HDBSCAN noise is treated as automatic disagreement, retains 6{,}186 FUAs (70.2\%). The 14.6 percentage-point gap quantifies the contribution of HDBSCAN's noise-reassignment step to the corroborated-core fraction. The core-set conclusion is qualitatively unchanged under either noise-handling rule: 70.2\% of FUAs agree with k-means via either non-noise HDBSCAN or GMM even when every HDBSCAN noise label is treated as automatic disagreement.

\subsection{Cross-method apparatus: HDBSCAN, GMM, Hungarian matching}
\label{sec:stability_apparatus}

The core-set construction in Methods~\ref{sec:stability} relies on two alternative clustering methods applied to the same PCA-reduced trajectories and matched against the k-means reference partition. We document the hyperparameter settings, the treatment of HDBSCAN's noise labels, the exclusion of spectral clustering, and the matching procedure here.

\textbf{HDBSCAN} \citep{Campello2013} was run with \texttt{min\_cluster\_size}=50 and \texttt{min\_samples}=15 under the default Euclidean metric. Under these settings HDBSCAN labels $3{,}754$ of the $8{,}808$ FUAs as noise. For the Hungarian-matched comparison with k-means these noise-labelled FUAs are reassigned to the nearest k-means centroid before matching; the alternative treatment, in which HDBSCAN noise is taken as automatic disagreement, is audited in Supplementary~\ref{sec:hdbscan_noise_audit}.

\textbf{Gaussian Mixture Model} was fitted with full-covariance components, \texttt{n\_components}=17, and \texttt{n\_init}=10, using the scikit-learn implementation \citep{Pedregosa2011}.

\textbf{Hungarian matching} \citep{Munkres1957} aligns cluster labels between each alternative partition and the k-means reference before per-regime Jaccard indices are computed. The matching maximises the trace of the cluster-cluster overlap matrix, ensuring each k-means regime is compared against its best-matching counterpart in the alternative partition.

\textbf{K-means restarts.} K-means uses \texttt{n\_init}=10 k-means++ restarts per seed during the silhouette sweep over $k = 3, \ldots, 20$, and \texttt{n\_init}=20 for the final fit at $k^* = 17$. The asymmetry reduces computation in the sweep without materially shifting $k^*$, since the silhouette curve at $k = 17$ exceeds its neighbours by more than the seed-to-seed standard deviation.

\subsection{Per-regime cross-method Jaccard indices}
\label{sec:jaccard_table}

To support the core-set construction in Methods \ref{sec:stability}, we report the per-regime Jaccard agreement between the k-means partition at $k^* = 17$ and each Hungarian-matched alternative-method partition (Table~\ref{tab:jaccard}). The two main-text alternatives are HDBSCAN and Gaussian Mixture Models; we additionally include Ward agglomerative clustering as a third independent method to confirm that the cross-method-stable regime set does not depend on the two-method choice.

\begin{table}[h]
    \centering
    \small
    \renewcommand{\arraystretch}{1.05}
    \caption{\textbf{Per-regime Jaccard indices against k-means at $k^* = 17$.} Each alternative method is Hungarian-matched to the k-means reference partition; Jaccard is computed per regime against its matched counterpart. Bold rows are \emph{cross-method-stable}: Jaccard $> 0.70$ on \emph{all three} alternatives. Numerical values are rounded to three decimal places.}
    \label{tab:jaccard}
    \begin{tabular}{lrrrc}
        \toprule
        \textbf{Regime} & \textbf{HDBSCAN} & \textbf{GMM} & \textbf{Ward} & \textbf{Stable?} \\
        \midrule
        C0                                & 0.559 & 0.667 & 0.805 & --- \\
        \textbf{C1} \textit{(China)}      & \textbf{0.804} & \textbf{0.751} & \textbf{0.883} & \checkmark \\
        C2                                & 0.582 & 0.674 & 0.952 & --- \\
        C3                                & 0.235 & 0.626 & 0.850 & --- \\
        \textbf{C4} \textit{(Venezuela)}  & \textbf{1.000} & \textbf{1.000} & \textbf{1.000} & \checkmark \\
        C5                                & 0.903 & 0.589 & 0.917 & --- \\
        C6                                & 0.706 & 0.593 & 0.733 & --- \\
        \textbf{C7} \textit{(Iraq)}       & \textbf{0.985} & \textbf{1.000} & \textbf{0.985} & \checkmark \\
        C8\textsuperscript{$\dagger$}     & 0.007 & 0.006 & 0.000 & --- \\
        C9                                & 0.902 & 0.545 & 0.908 & --- \\
        C10                               & 0.271 & 0.302 & 0.537 & --- \\
        C11                               & 0.107 & 0.103 & 0.221 & --- \\
        C12                               & 0.019 & 0.026 & 0.042 & --- \\
        \textbf{C13} \textit{(Iran)}      & \textbf{0.878} & \textbf{0.973} & \textbf{0.878} & \checkmark \\
        C14                               & 0.165 & 0.195 & 0.000 & --- \\
        C15                               & 0.889 & 0.617 & 0.938 & --- \\
        C16                               & 0.821 & 0.505 & 0.463 & --- \\
        \bottomrule
    \end{tabular}

    \vspace{4pt}
    \footnotesize
    \raggedright
    $^{\dagger}$ Cluster 8 contains a single FUA and is reported for completeness; it does not contribute to regime-level analyses.
    \normalsize
\end{table}

\noindent The four cross-method-stable regimes (C1, C4, C7, C13) all comprise FUAs from a single dominant country (China, Venezuela, Iraq, Iran respectively), and each retains Jaccard $> 0.70$ even when Ward---a method we did not use to construct the core set---is added as a third reference. The remaining regimes survive the corroborated core-set criterion in Methods\ref{sec:stability} but have method-specific boundaries; they should be interpreted as regions of trajectory space rather than as sharply bounded partitions.

\paragraph{Threshold sensitivity.} The four-regime headline is robust to reasonable choices of the Jaccard threshold. The set $\{\text{C4}, \text{C7}, \text{C13}\}$ passes any threshold in $[0.50, 1.00]$ on all three methods; C1 passes at $0.70$ but not at $0.80$ (constrained by GMM $= 0.751$). Lowering the threshold to $0.60$ admits C15 as a fifth stable regime (HDBSCAN $0.889$, GMM $0.617$, Ward $0.938$); raising it to $0.80$ drops C1, leaving three regimes. We adopt the $0.70$ threshold following the partition-comparison literature, where it is conventionally used as a strong-agreement cut-point.

\subsection{Bootstrap stability (FUA resampling)}
\label{sec:bootstrap_jaccard}

To complement the cross-method Jaccard test of Section~\ref{sec:jaccard_table}, we assess each regime's stability under FUA-level resampling using the bootstrap-Jaccard procedure of \citet{Hennig2007}. We draw $B = 100$ bootstrap samples of the $N = 8{,}808$ FUAs (with replacement), re-fit the full preprocessing-plus-k-means pipeline at $k^* = 17$ on each, Hungarian-match the resulting partition to the reference partition, and compute the per-regime Jaccard against the matched bootstrap counterpart. We report the mean and median Jaccard across $B$, the 95\% percentile interval, and the fraction of bootstraps achieving Jaccard $> 0.70$ (Table~\ref{tab:bootstrap_jaccard}). The choice of $B = 100$ sits at the lower end of \citeauthor{Hennig2007}'s recommended range; for the four cross-method-stable regimes the mean Jaccard is stable across resamples (the Monte Carlo standard error on the mean is below $0.02$), while the wider 95\% intervals reported for smaller regimes should be read as lower-bound estimates of bootstrap uncertainty.

\begin{table}[h]
    \centering
    \small
    \renewcommand{\arraystretch}{1.05}
    \caption{\textbf{Per-regime bootstrap-Jaccard stability ($B = 100$ FUA-bootstrap samples).} Mean, median, and 95\% percentile interval of the Jaccard index between the reference $k^* = 17$ partition and Hungarian-matched bootstrap partitions. ``\%${>}0.70$'' is the fraction of bootstrap iterations in which the regime achieves Jaccard $> 0.70$. Bold rows mark the four cross-method-stable regimes (Table~\ref{tab:jaccard}). The reserved-empty label $C_8$ is omitted (see Table~\ref{tab:cluster_stats} footnote).}
    \label{tab:bootstrap_jaccard}
    \begin{tabular}{lrrrrr}
        \toprule
        \textbf{Regime} & $n$ & \textbf{Mean} & \textbf{Median} & \textbf{95\% CI} & \textbf{\%${>}0.70$} \\
        \midrule
        C0  & 589  & 0.756 & 0.834 & [0.00, 0.94] & 74\% \\
        \textbf{C1} \textit{(China)}    & 1430 & \textbf{0.798} & 0.788 & [0.59, 0.93] & \textbf{84\%} \\
        C2  & 627  & 0.899 & 0.918 & [0.70, 0.97] & 96\% \\
        C3  & 1208 & 0.893 & 0.948 & [0.60, 0.98] & 94\% \\
        \textbf{C4} \textit{(Venezuela)}& 64   & \textbf{1.000} & 1.000 & [1.00, 1.00] & \textbf{100\%} \\
        C5  & 1880 & 0.920 & 0.943 & [0.80, 0.96] & 100\% \\
        C6  & 322  & 0.887 & 0.928 & [0.71, 0.98] & 100\% \\
        \textbf{C7} \textit{(Iraq)}     & 64   & \textbf{0.991} & 1.000 & [0.93, 1.00] & \textbf{100\%} \\
        C9  & 449  & 0.764 & 0.823 & [0.24, 0.96] & 60\% \\
        C10 & 1015 & 0.674 & 0.850 & [0.00, 0.92] & 70\% \\
        C11 & 129  & 0.119 & 0.014 & [0.00, 0.87] & 6\%  \\
        C12 & 24   & 0.434 & 0.344 & [0.00, 1.00] & 35\% \\
        \textbf{C13} \textit{(Iran)}    & 188  & \textbf{0.911} & 0.995 & [0.12, 1.00] & \textbf{94\%} \\
        C14 & 402  & 0.627 & 0.695 & [0.02, 0.90] & 47\% \\
        C15 & 142  & 0.487 & 0.726 & [0.00, 0.91] & 58\% \\
        C16 & 275  & 0.792 & 0.828 & [0.01, 0.91] & 93\% \\
        \bottomrule
    \end{tabular}
\end{table}

All four cross-method-stable regimes (C1, C4, C7, C13) retain bootstrap mean Jaccard $> 0.70$ under FUA-level resampling. Three of these four (C4, C7, C13) achieve $> 94\%$ of bootstrap iterations above the threshold; C1 sits at 84\%, well above the 50\% mark that would indicate borderline stability. The bootstrap test admits a broader stable set than the cross-method test: 11 of the 17 regimes pass bootstrap mean $> 0.70$, indicating that the cross-method criterion is the stricter of the two stability tests. Three regimes (C11, C12, C15) fail bootstrap stability decisively (mean $< 0.5$), consistent with their interpretation as method-specific zones of trajectory space rather than sharply bounded regimes.

The wide percentile interval on C13 (lower bound $0.12$) reflects a small minority of bootstrap iterations producing very low Jaccard for this regime, consistent with C13's small size ($n = 188$) and its known fragility under joint method-and-$k$ perturbation (Section~\ref{sec:supercore}). The median ($0.995$) and high \%${>}0.70$ rate are the primary bootstrap-Jaccard statistics for this regime; we report the median because it is less sensitive to the small number of bootstrap iterations in which Hungarian matching fails for the smallest regimes.

\subsection{Cross-$k$ sensitivity}
\label{sec:cross_k}

We report two complementary views of how the $k = 17$ partition responds to changes in $k$.

\textbf{S4a: stability metrics versus $k$.} Across $k \in \{15, \ldots, 19\}$, the $k = 17$ partition sits at or near the silhouette peak, matched-cluster Adjusted Rand Index between $k = 17$ and each alternative $k$ exceeds 0.85, and mean per-regime Jaccard exceeds 0.70. The partition is therefore stable under small perturbations of the cluster count.

\begin{figure}[h]
    \centering
    \includegraphics[width=0.78\linewidth]{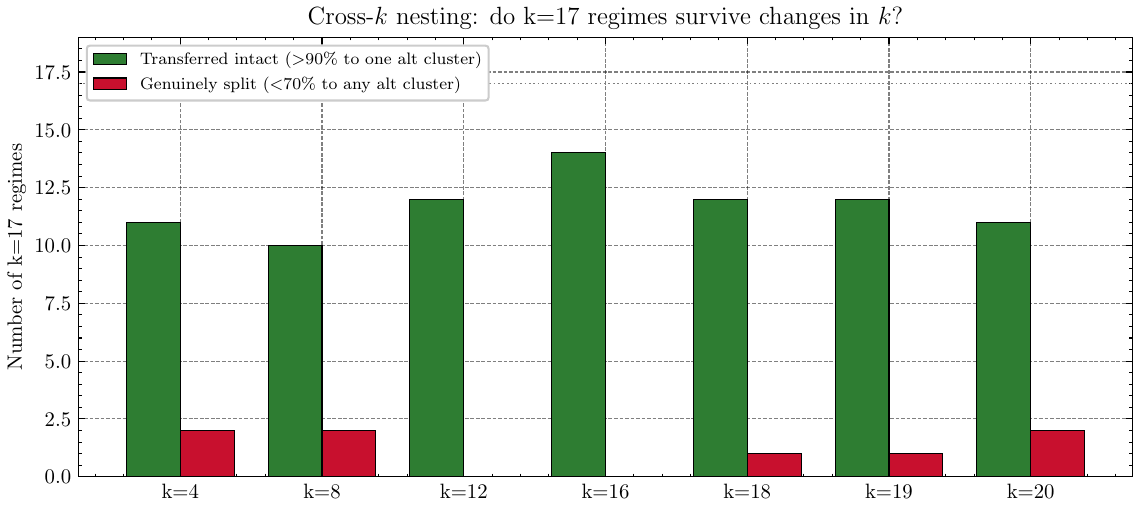}
    \caption{\textbf{Regime nesting under $k$ variation.} Each row shows one alternative $k$; coloured cells indicate whether each $k = 17$ regime transfers intact, splits, or merges with another. At $k = 16$, 14 of 17 regimes transfer intact and none split; at $k = 18$, 12 transfer intact and 1 splits; at $k = 20$, 11 transfer intact and 2 split. Extreme values ($k = 4$, $k = 8$) dissolve the structure by forced merging.}
    \label{fig:cross_k_nesting}
\end{figure}

\textbf{S4b: regime nesting at neighbouring $k$.} For each alternative $k$, each of the 17 k-means regimes at $k = 17$ is classified as \emph{transferred intact} ($> 90\%$ of members map to one alternative-$k$ cluster) or \emph{genuinely split} ($< 70\%$ map to any single alternative cluster). Figure~\ref{fig:cross_k_nesting} reports the result.

The $k = 17$ partition does not globally reorganise as $k$ varies within $\pm 3$---it refines. This nesting property is the basis for the pairwise super-core robustness measure in Section \ref{sec:supercore}, which formalises ``which FUAs persist together across all these variations?"

\subsection{Pairwise super-core: robustness across method and $k$}
\label{sec:supercore}

\begin{figure}[h]
    \centering
    \includegraphics[width=0.92\linewidth]{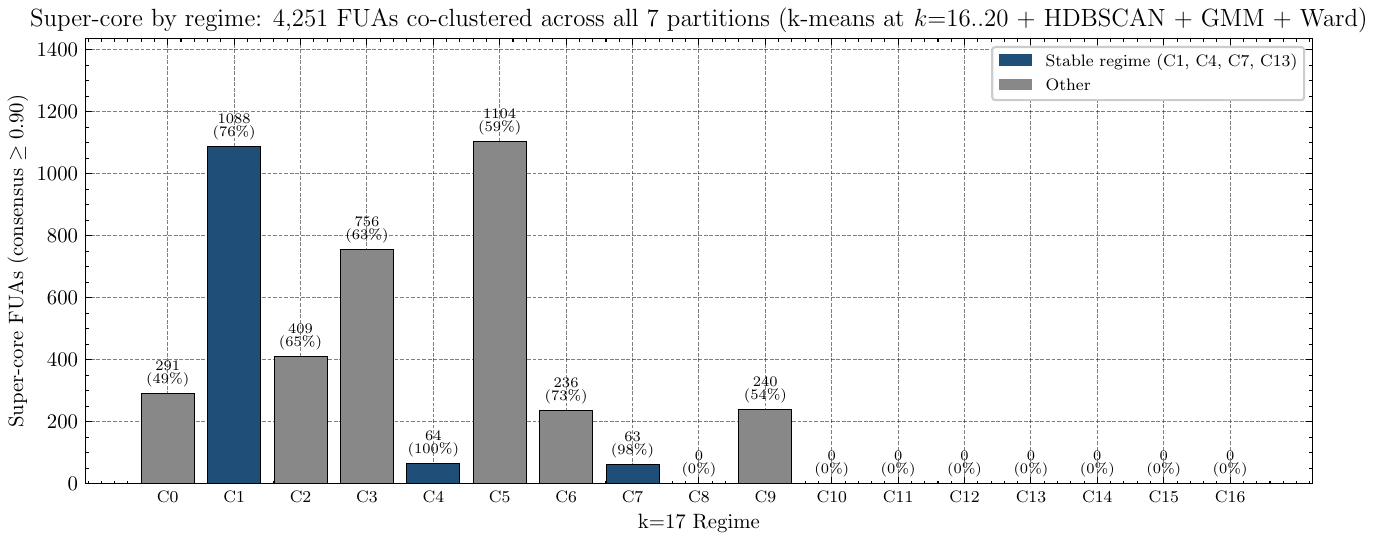}
    \caption{\textbf{Super-core fraction by regime.} For each k-means regime at $k=17$, we report the fraction of FUAs whose mean pairwise co-assignment across 8 alternative partitions exceeds $0.90$. Highest super-core share: C4 Venezuela ($100\%$), C7 Iraq ($98\%$), C1 China-dominated ($76\%$), C6 Russia ($73\%$).}
    \label{fig:supercore}
\end{figure}

Methods S\ref{sec:stability} tests robustness to method choice at fixed $k=17$, and Section \ref{sec:cross_k} reports robustness to $k$. We combine both into a single label-free measure. For each FUA $i$, we compute a \emph{consensus score} as the fraction of $i$'s same-cluster neighbours in the $k=17$ k-means partition that are also same-cluster neighbours in each of seven alternative partitions: k-means at $k \in \{16, 18, 19, 20\}$ and HDBSCAN, GMM, and Ward at $k=17$. We average across the seven, and define the \emph{super-core} as the set of FUAs whose mean pairwise co-assignment is $\geq 0.90$ (Figure.~\ref{fig:supercore}).

This measure is label-free: it does not rely on Hungarian matching, and is therefore immune to label-correspondence artefacts. It collapses both method-choice and $k$-choice variation into one score per FUA.

\textbf{Result.} 4{,}251 FUAs ($48.3\%$) form the super-core. The super-core fraction by regime ranges from $100\%$ (C4 Venezuela) to $0\%$ (C10, C11, C12, C13, C14, C15, and C16). The four regimes with the highest super-core share are C4 Venezuela ($100\%$), C7 Iraq ($98\%$), C1 China-dominated ($76\%$), and C6 Russia ($73\%$). The fixed-$k$ cross-method-stable set, by contrast, is \{C1, C4, C7, C13\} (Jaccard $>0.70$ on both HDBSCAN and GMM; see Section~\ref{sec:stability}). The two sets share three regimes (C1, C4, and C7), which are robust by both criteria; C6 and C13 each pass one criterion but fail the other. The fixed-$k$ stability and label-free super-core criteria therefore agree on the strongest core (C1, C4, C7) but diverge on the marginal cases, which we unpack next.

\textbf{C13 Iran and C6 Russia: criterion-specific results.} C13 (Iran) passes the cross-method Jaccard threshold at $k=17$ (HDBSCAN $0.878$, GMM $0.973$; Table~\ref{tab:jaccard}) but has zero super-core FUAs because, at $k=18$, the Iran regime is absorbed into a larger cluster, diluting pairwise co-assignment across $k$. C6 (Russia / former-Soviet bloc) shows the mirror pattern: it has a $73\%$ super-core share but its per-regime Jaccard against GMM is $0.593$ (Table~\ref{tab:jaccard}), below the $0.70$ threshold, so it does not appear in the fixed-$k$ stable set. Cross-method stability (at fixed $k$) and cross-$k$ stability are therefore distinct robustness notions. Cross-method robustness rewards label-level agreement under a single partitioning, whereas the super-core rewards label-free neighbourhood preservation under joint method and $k$ perturbation. The asymmetry between C13 and C6 reflects this distinction directly.

\subsection{Alternative trajectory-shape clustering (across-time standardisation)}
\label{app:rowwise}
This section establishes that the regime structure recovered in the main analysis reflects relative-rank trajectories rather than abstract trajectory shape, and is therefore not invariant to the choice of standardisation. The main analysis standardises growth rates cross-sectionally (per year, $z$-scored across the FUA panel), preserving how each FUA ranks against its peers in each year. An alternative -- standardising each FUA's trajectory across time, so that each FUA's mean and scale are erased -- clusters on trajectory shape alone (V-shaped recovery, smooth growth, volatile, flat) independent of relative position. The two standardisations encode different research questions: cross-sectional clustering identifies FUAs that occupy similar positions in their year-cohort over time, while across-time clustering identifies FUAs that follow similar trajectory shapes regardless of cohort context.

We ran the full clustering pipeline (PCA at 80\% variance, k-means $k = 17$, cross-method stability assessment) under across-time standardisation as a sensitivity check. The resulting partition has ARI $= 0.39$ against the main partition, indicating substantially different clusters. Across-time preprocessing yields $d_{\mathrm{PCA}} = 11$ components (versus $d = 14$ in the main analysis), and spectral clustering, which is degenerate in the main analysis (matched-cluster ARI $\approx 0.001$ against k-means), recovers a sensible partition here (matched-cluster ARI $\approx 0.48$). We therefore extend the core-set criterion of Section~\ref{sec:stability} to three non-degenerate alternatives (HDBSCAN, GMM, spectral), requiring agreement with at least two of three; this yields $3{,}013$ FUAs ($34.2\%$), compared with $7{,}474$ ($84.8\%$) in the main analysis with its two-method criterion. No regime passes Jaccard $> 0.70$ on any method combination under across-time preprocessing.

The low cross-standardisation agreement is informative beyond a sensitivity check: it indicates that the main-analysis regimes are not artefacts of any FUA's individual trajectory shape, but emerge from the comparative position FUAs occupy relative to their peers in each year. This pattern is consistent with the interpretation, advanced in the main text, that regimes reflect structural features governing relative rather than absolute growth dynamics. The cross-sectional standardisation chosen for the main analysis encodes this conception explicitly; the across-time analysis confirms that a shape-typological alternative fails to recover comparable structure.

\end{document}